\documentclass[lettersize,journal]{IEEEtran}
\usepackage{amsmath,amsfonts}
\usepackage{algorithmic}
\usepackage{algorithm}
\usepackage{array}
\usepackage[caption=false,font=normalsize,labelfont=sf,textfont=sf]{subfig}
\usepackage[table]{xcolor}
\usepackage{textcomp}
\usepackage{amsmath}
\usepackage{stfloats}
\usepackage{url}
\usepackage{booktabs}
\usepackage{verbatim}
\usepackage{graphicx}
\usepackage{cite}
\usepackage{listings}
\usepackage{amsmath,amsfonts,color}
\usepackage{graphicx}
\usepackage{amsfonts}
\usepackage{array}
\usepackage{amssymb}
\usepackage{helvet}
\usepackage{color}
\usepackage{bm}
\usepackage{float}
\usepackage{cite}
\usepackage{multirow}
\usepackage[caption=false,font=normalsize,labelfont=sf,textfont=sf]{subfig}
\usepackage{amsmath,amsfonts}
\usepackage{algorithmic}
\usepackage{algorithm}
\usepackage{textcomp}
\usepackage{stfloats}
\usepackage{url}
\usepackage{verbatim}
\usepackage{graphicx}
\usepackage{cite}
\usepackage{amssymb}
\usepackage{makecell}
\usepackage{mathrsfs}
\usepackage{url}
\usepackage{xurl}
\usepackage[hidelinks]{hyperref}

\newcommand{\RNum}[1]{\uppercase\expandafter{\romannumeral #1\relax}}

\begin{document}

\title{\textcolor{black}{Networked ISAC Enabled Target Recognition \\ Towards Low-Altitude Economy}}
\author{\textcolor{black}{Hongliang Luo, Chuanbin Zhao, Boxuan Sun, Zhonghua Chu, Shengjie Quan, Guangyi Liu,  \\ and Feifei Gao,~\IEEEmembership{Fellow,~IEEE}}
\thanks{H. Luo, C. Zhao, B. Sun, Z. Chu, S. Quan,  and F. Gao are with Department of Automation, Tsinghua University, Beijing 100084, China (email: luohl23@mails.tsinghua.edu.cn;
zcb23@mails.tsinghua.edu.cn;
sunbx26@163.com;
	 c3zhua@163.com;
	 quansj23@mails.tsinghua.edu.cn;   feifeigao@ieee.org).
	 
	 G. Liu is with China Mobile Research Institute, Beijing 100053,
	 China (e-mail: liuguangyi@chinamobile.com).
}
}



\maketitle

\begin{abstract}
In this paper,
we propose a \textcolor{black}{\emph{low-altitude  target (LAT)}} recognition scheme based on multi-base station (BS) collaboration and multi-scale feature fusion 
for integrated sensing and communications (ISAC) network. 
\textcolor{black}{Firstly,}
we formulate the motion equations, echo channels,  and echo signals for unmanned aerial vehicle (UAV), bird, 
vehicle, and pedestrian under multi-BS  collaborative monitoring scenario.
Then we extract the 
velocity-resolution-preferred time-frequency spectrum,  time-resolution-preferred time-frequency spectrum, and the velocity-transfer time-frequency spectrum observed by each BS from  echo signals.
We collectively refer to these three types of time-frequency spectrum as the \emph{multi-scale feature} of the LAT. 
Next, we design a multi-BS and multi-scale feature fusion enabled LAT recognition network with Swin Transformer, 
which 
employs the visualized images of multi-scale feature
to jointly recognize the target through
deep feature extraction, intra-BS feature interaction, inter-BS feature interaction, and target recognition output.
We generate a massive echo signal dataset comprising 1,440,000 samples for LAT recognition within ISAC network. This dataset can serve as a public benchmark to evaluate our proposed scheme and facilitate future research.
Simulation results demonstrate that the proposed scheme realizes high recognition accuracy and robust unseen-subtype generalization, confirming the effectiveness of multi-scale feature fusion and the additional gains brought by multi-BS collaboration.
The project page is available at:
\url{https://alivn999.github.io/COSMOS-Networked-ISAC-Enabled-Target-Recognition-Towards-Low-Altitude-Economy/}.
\end{abstract}


\section{Introduction}

\textcolor{black}{Low-altitude economy} (LAE) is emerging as a significant direction for global technological innovation and economic development\cite{10723207,10608169,HUANG2024100694,2023arXiv231109047J}.
The core elements of LAE are unmanned aerial vehicle (UAV) and other general aviation aircraft\cite{XIANG2024100140},  giving rise to plentiful flight  activities such as low-altitude logistics,  inspections,  travel, surveying, firefighting\cite{8795473,9666755}.
\textcolor{black}{To support  the operation of LAE,}
  extensive physical infrastructures and information infrastructures are required. 
The former encompasses take-off and landing sites, charging stations,  etc., while the latter comprises low-altitude communications, sensing, navigation, management networks, etc.

\textcolor{black}{However,}
the rapid proliferation of UAVs has inevitably led to many unauthorized and unregulated flight behaviors, posing significant threats to public safety and privacy. Consequently, it is imperative to leverage advanced information technologies to simultaneously monitor both cooperative UAVs and non-cooperative UAVs, thereby effectively ensuring low-altitude security\cite{drones3010013,8675384,10077453}.
Meanwhile, the emerging sixth generation mobile communications (6G) network is expected to integrate wireless communications and radar sensing functionalities into the same base station (BS), thereby establishing the  integrated sensing and communications (ISAC) system.
The ISAC system, equipped with active detection capabilities, can effectively monitor both cooperative UAVs and non-cooperative UAVs in low-altitude scenarios\cite{202310141,11250835,9040264,9606831,itu}.

Research on ISAC-enabled low-altitude monitoring technologies primarily focus on clutter suppression, target detection, parameter estimation, trajectory management, etc.
For example, R.~Zhang~\emph{et~al.}  propose a Go decomposition based clutter suppression and target detection algorithm for monostatic  ISAC system\cite{11162253}. 
X.~Lu~\emph{et~al.} apply multiple signal classification (MUSIC)  algorithm  to estimate the angle, distance and velocity of  UAV target in monostatic ISAC system~\cite{2024arXiv240412705L}.
S.~Yan~\emph{et~al.} propose a UAV trajectory monitoring scheme based on monostatic  ISAC system, which enables the real-time discovery of new UAV targets and the continuous tracking of  discovered UAV targets\cite{11134125}.  
To further enhance the monitoring accuracy and robustness, researchers explore multi-BS cooperative sensing technologies within ISAC network. For example,
Z.~Wei~\emph{et~al.} propose a symbol-level multi-BSs cooperative sensing scheme based on fast Fourier transform (FFT) data compression and lattice search\cite{10226276}.
Y.~Huang~\emph{et~al.} formulate  multi-BS cooperative localization  as a sparse imaging problem based on compressive sensing, thereby enhancing the system's capability to detect sparse UAV targets across expansive low-altitude airspace\cite{11151696}.
These works provide comprehensive technical preparations and valuable insights for the localization and tracking of UAV targets.

However, \textcolor{black}{an easily overlooked problem} is that ISAC system suffers from severe false alarms when performing low-altitude monitoring. These false alarms mainly come from two aspects.
On the one hand,
when the BS transmits sensing detection signals to low-altitude area through beamforming, the birds would also be illuminated by the main-lobe of the transmitted beam and thus  cause echoes.
On the other hand, ISAC system operating in  sub-6 GHz band is typically equipped with smaller-scale antenna arrays, particularly with only a few antenna elements in the vertical dimension.  Hence the beam in the vertical dimension possesses a large side-lobe, which will illuminate some vehicles or pedestrians on the ground, thereby causing echoes.
Although the array gain of beam side-lobe is significantly lower than that of the main-lobe, the radar cross section (RCS) of vehicles and pedestrians is much larger than that of UAVs. Consequently, these vehicles and pedestrians still induce severe false alarms for UAV detection and monitor.

To mitigate these false alarms triggered by non-UAV targets, a natural approach is to distinguish UAVs from birds, vehicles, and pedestrians based on the echo signals.
This sensing task \textcolor{black}{is generally}  referred to as \emph{LAT recognition}.
For example, J.~Wei~\emph{et~al.}  propose a
rotor micro-Doppler null space pursuit  algorithm to obtain the specific feature caused by UAV's paddle rotation, and 
the sinusoidal envelope characteristics exhibited in the time-Doppler spectrum are expected to be  used for the  recognition of rotary UAV\cite{11077832}. 
D.~Ma~\emph{et~al.}   evaluate the performance of UAV identification using various time-division duplex (TDD) patterns under 
the fifth generation mobile communications (5G) network\cite{10625724}.
J.~Xue~\emph{et~al.}  utilize the Short-Time Fourier Transform (STFT) to obtain the micro-Doppler spectral images of UAVs and birds from the echo signals, and then employ the DC-Former network to identify UAVs and  birds\cite{11069481}.
Besides, our prior work \cite{11159257} proposes a UAV and bird recognition scheme for millimeter-wave ISAC system by fusing micro-Doppler spectrum with high resolution range profile (HRRP), in which  a  dataset within monostatic  ISAC system containing over 230,000 echo signal samples is constructed to train, validate, and test that scheme.

Nevertheless, 
at present, ISAC-enabled \textcolor{black}{LAT} recognition remains at its early stage, and there are still many important technical issues that require further investigation.
\textcolor{black}{For example,}  existing \textcolor{black}{LAT} recognition networks \cite{11069481,11159257} primarily focus on UAVs and birds, while neglecting the equally troublesome presence of vehicles and pedestrians. 
Besides,  \cite{11077832,10625724,11069481,11159257} typically employ microsecond-level symbol intervals when extracting the  micro-Doppler spectrum of target. However, the symbol intervals in practical sub-6 GHz ISAC systems for target recognition are usually at the millisecond-level\cite{RN16}, which significantly degrades the resolution of time-frequency analysis. Under such circumstances, it becomes imperative to consider how to enhance target recognition accuracy within the constraints of limited time-domain resources configuration.
Moreover, many critical application scenarios impose stringent requirements on the false alarm rate of low-altitude monitoring system.  Hence it is essential to further improve the accuracy of \textcolor{black}{LAT} recognition, with potential pathways including multi-feature fusion, multi-BS cooperation,  multi-modal fusion, etc.

Motivated by the aforementioned observations,
we propose a \textcolor{black}{LAT} recognition scheme based on multi-BS collaboration and multi-scale feature fusion 
for ISAC network.  
The contributions of this paper are summarized as follows.

\begin{itemize}
	
\item  We formulate the motion processes,
  echo channels, and echo signals for 
  UAVs, birds, vehicles and pedestrians
  under multi-BS collaborative sensing scenario.

\item We propose a multi-scale time-frequency feature extraction algorithm for ISAC system.
Specifically,  we utilize STFT with  two sets of parameters to extract
the velocity-resolution-preferred time-frequency spectrum (VRP-TF spectrum), and  the  time-resolution-preferred time-frequency spectrum (TRP-TF spectrum) of the target from the echo signals.
Meanwhile,  we utilize matching filtering and differential calculation to extract 
the velocity-transfer time-frequency spectrum (VT-TF spectrum) of the target from the echo signals.
The  VRP-TF spectrum, TRP-TF spectrum, and VT-TF spectrum  are  referred to as
 the \emph{multi-scale feature} of the \textcolor{black}{LAT}.

\item We design a multi-BS and multi-scale feature fusion enabled \textcolor{black}{LAT} recognition network with Swin Transformer, which 
employs the visualized images of multi-scale feature
to jointly recognize the target through
deep feature extraction, intra-BS feature interaction, inter-BS feature interaction, and target recognition output.

\item  We generate a massive echo signal dataset comprising 1,440,000 samples for LAT recognition within ISAC network. This dataset can serve as a public benchmark to evaluate the proposed scheme and facilitate future research.
The effectiveness of the  proposed    scheme   has been demonstrated by simulation results.

\end{itemize}

The remainder of this paper is organized as follows.
In Section \RNum{2}, we  construct 
the multi-BS cooperative sensing scenario, as well as  the echo channel model of \textcolor{black}{LAT}. 
In Section \RNum{3}, we \textcolor{black}{formulate} the echo signals of  UAV, bird, vehicle and pedestrian   under   ISAC network.
In Section~\RNum{4}, we propose the
multi-BS and multi-scale feature fusion enabled 
\textcolor{black}{LAT} recognition scheme. 
Simulation results and conclusions are given in Section~\RNum{5} and Section~\RNum{6}.

\begin{figure}[!t]
	\centering
	\includegraphics[width=85mm]{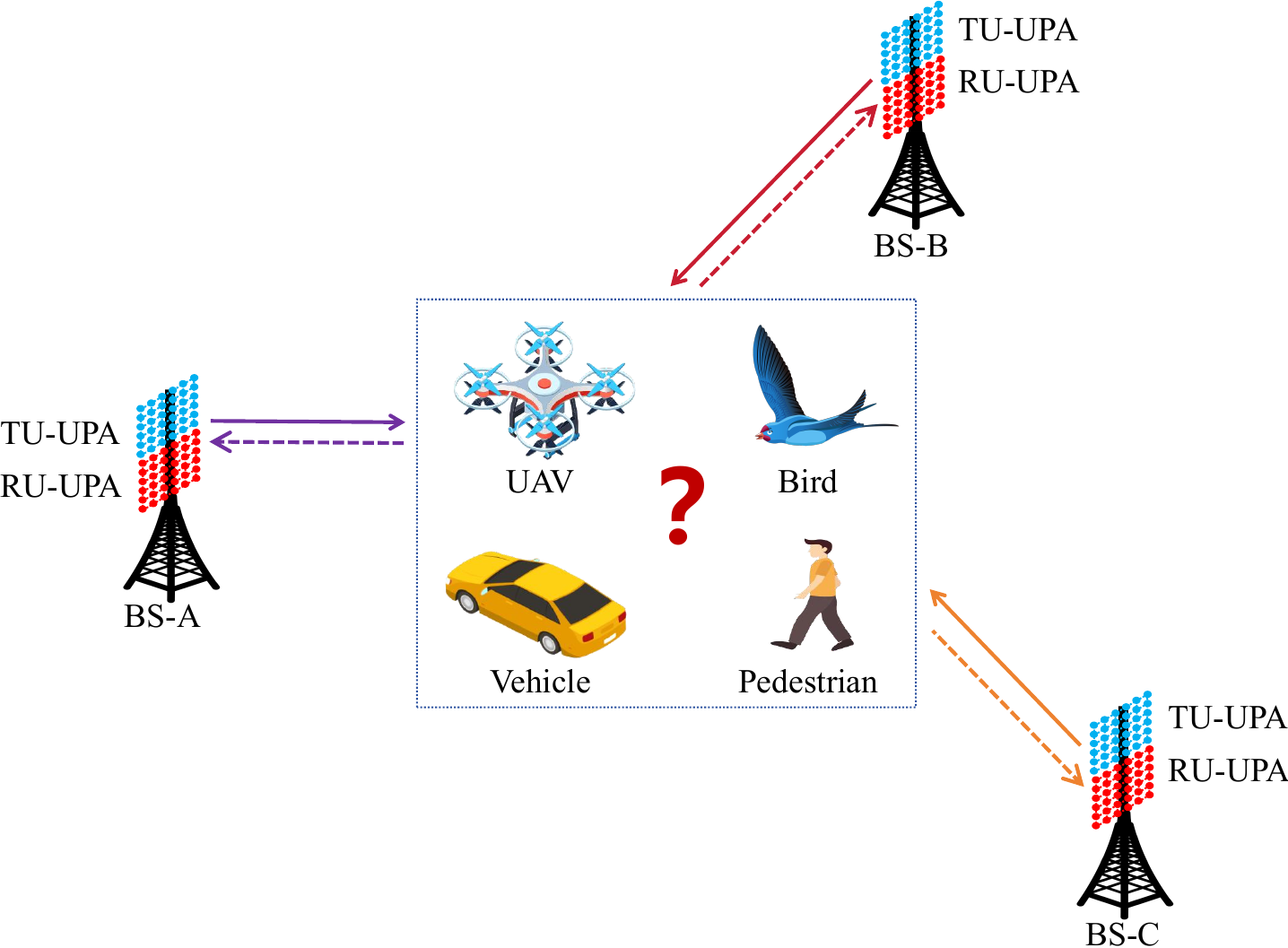}
	\caption{LAT recognition scenario  in ISAC network.}
	\label{fig_1}
\end{figure}

\emph{Notation}:
Lower-case and upper-case boldface letters $\mathbf{a}$ and $\mathbf{A}$ denote a vector and a matrix; $\mathbf{a}^*$, 
$\mathbf{a}^T$ and $\mathbf{a}^H$ denote the conjugate,  transpose and  conjugate transpose of  $\mathbf{a}$, respectively;
$\mathbf{a}[n]$  denotes the $n$-th element of the vector $\mathbf{a}$;
$\mathbf{A}[{i,j}]$ denotes the $(i,j)$-th element of the matrix $\mathbf{A}$; $\mathbf{A}[i_1:i_2,:]$ is the submatrix composed of all columns elements in rows $i_1$ to $i_2$ of matrix $\mathbf{A}$;
$\mathbf{A}[:,j_1:j_2]$ is the submatrix composed of all rows elements in columns $j_1$ to $j_2$ of matrix $\mathbf{A}$;
$\mathbb{S}$ denotes a set; 
${\rm card}(\mathbb{S})$ represents the number of elements in set $\mathbb{S}$;
$\varnothing$ denotes an empty set; 
$\cap$ and $\cup$ respectively represent the intersection and union of sets;
$||\cdot||_2$  represents the Euclidean norm;
$\lfloor \cdot \rfloor$ represents the floor  function.

\section{ISAC Scenario and  System Model}

In this section, 
we derive the echo channel model for the \textcolor{black}{LAT} under multi-BS collaborative sensing scenario.

\subsection{Multi-BS Cooperative Sensing Scenario and System Model}

Fig.~1 depicts the
\textcolor{black}{LAT} recognition scenario with multi-BS cooperation in ISAC network, in which three BSs, namely BS-A, BS-B, and BS-C, constitute a basic sensing cell. Meanwhile, there are four categories of  \textcolor{black}{LATs}: UAV, bird, vehicle, and pedestrian.
 Generally, the BSs can transmit sensing detection signals, receive and process the sensing  echo signals to obtain various information of the  \textcolor{black}{LATs},
 thereby accomplishing various sensing tasks.
Here we  focus on how to use ISAC network to realize   high-precision \textcolor{black}{LAT} recognition.
 

\begin{figure}[!t]
	\centering
	\includegraphics[width=90mm]{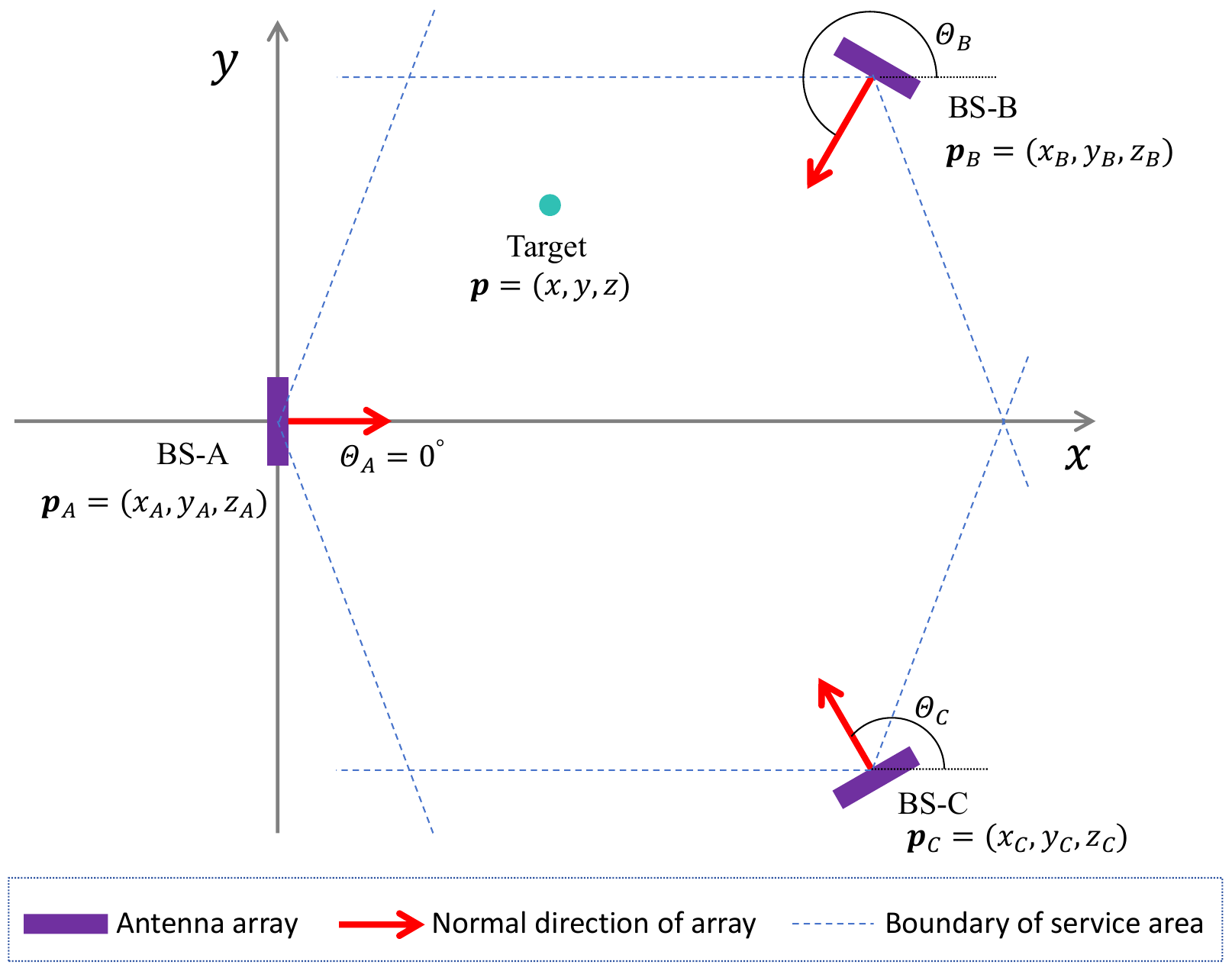}
	\caption{Aerial view of multi-BS cooperative sensing scenario.}
	\label{fig_1}
\end{figure}

\textcolor{black}{Each}  BS employs massive multiple input multiple output (MIMO) arrays and orthogonal frequency division multiplexing (OFDM) signals,
\textcolor{black}{and  operates} in  sub-6 GHz frequency bands to realize both  wireless communications and low-altitude  monitoring.
As shown in Fig.~1, each BS is equipped with one {transmitting unit uniform planar array (TU-UPA)} and one {receiving   unit uniform planar array (RU-UPA)}.
Both TU-UPA and RU-UPA  are perpendicular to the horizontal plane.
TU-UPA and  RU-UPA in the same BS 
are co-located  and are parallel to each other,  such that they can see the  low-altitude  targets at the same propagation directions\cite{9898900}. 
The 
TU-UPA and RU-UPA of the $k$-th BS are each equipped with $N_{T,k}=N^{\parallel}_{T,k}\times N_{T,k}^{\perp}$ and $N_{R,k}=N^{\parallel}_{R,k}\times N_{R,k}^{\perp}$ antenna elements,
where $k \in \{A,B,C\}$, $\parallel$  represents the direction parallel to the horizontal plane, and $\perp$  represents the direction perpendicular to the horizontal plane.  The antenna spacing between the antennas  along horizontal direction and vertical direction 
 are 
$d_{\parallel} = d_{\perp} =  d = \frac{\lambda}{2}$,
where $\lambda = c/f_0$ is the wavelength, $f_0$ is   the operating frequency of the ISAC network, and $c$ represents the speed of light.

 Fig.~2 illustrates \textcolor{black}{the} geometric configurations of the BSs from an aerial view. 
 \textcolor{black}{Let us  establish a global coordinate system $O$-$xyz$ to facilitate the representation of spatial positions.}
 The Cartesian coordinate position of the $k$-th BS in  global coordinate system is 
 $\mathbf{p}_k = [x_k,y_k,z_k]^T$, where $k \in \{A,B,C\}$. 
Without loss of generality, we assume that the location of BS-A satisfies $x_A = y_A = 0$.
 Since the UPAs of all BSs are perpendicular to  horizontal plane, the angle between the normal direction of each UPA and the XOY plane is $0^\circ$. 
The angle between
the projection vector of the $k$-th BS's UPA normal vector onto the XOY plane and 
 the positive direction of  $x$-axis 
   is $\Theta_{k}$, where $k \in \{A,B,C\}$,
   and $0^\circ \leq \Theta_{k} < 360^\circ $.

In terms of frequency domain resource allocation, 
to avoid mutual interference among multiple BSs, we allocate different sub-frequency band to each BS,
\textcolor{black}{as did in  \cite{10226276}.}
Assuming the subcarrier  interval of  OFDM signal is $\Delta f$, and each BS utilizes $M$ subcarriers. Then the sensing transmission bandwidth of each BS is $W = M \Delta f$.
Let us denote the lowest frequency subcarriers of BS-A, BS-B, and BS-C as $f_A$, $f_B$, and $f_C$, respectively. The $m$-th subcarrier frequency of the $k$-th BS is $f_{k,m} \!=\! f_k + m\Delta f$,
$k \!\in\! \{A,B,C\}$, $m=0,1,...,M\!-\!1$.
Assuming a protection interval of $M_{\rm protect}$  subcarriers is enabled between adjacent BSs.
Then we can arrange $f_A = f_0$, $f_B = f_{A,M-1}+M_{\rm protect}\Delta f$, and $f_C = f_{B,M-1}+M_{\rm protect}\Delta f$ to avoid mutual interference.


In terms of time domain resource allocation, 
 let us represent the  time interval between  adjacent OFDM symbols as  $T_s$. 
According to  5G new radio (NR) standard, the basic time unit $T_s$ is at the microsecond level. 
However, due to the limited time-frequency resources in  sub-6 GHz band and the necessity to support high-speed communications services for numerous users, it is impractical for actual ISAC system to allocate thousands of consecutive symbol units, spaced by $T_s$, for the single LAT recognition task.
Consequently, the practical BS system adopts a sparse sampling strategy by selecting one symbol out of every $P$ symbols for target  recognition purposes, while other symbols are used for communications tasks and other sensing tasks\cite{RN16}.
These uniformly selected symbols for target recognition are aggregated into a new \emph{target recognition frame (TRF)}, which spans multiple radio frames as defined in the 5G NR standard.
For clarity, we directly consider that one \textcolor{black}{TRF} contains $N$  OFDM symbols, and 
the time interval between two adjacent symbols
in the constructed TRF is $T_r$, which satisfies $T_r = P T_s$.

\subsection{
	Echo Channel Model of  LAT in ISAC Network}

Since  BS operates in a distinct sub-frequency band,  each BS \textcolor{black}{would transmit} the sensing detection signals via its own TU-UPA. These sensing detection signals are subsequently scattered by  LAT, and the resulting echo signals are captured by the corresponding RU-UPA at each  BS for further processing.
\textcolor{black}{
Following the methodology in \cite{11159257}, we adopt a multi-scattering-point representation for LAT, where the echo channel of  LAT is approximated by the superposition of the channels associated with individual scattering points~\cite{9947033}. 
It should be emphasized that this representation is not intended to imply that all scattering points can be individually resolved by a practical sub-6 GHz ISAC BS. 
Instead, the point-cloud model provides a tractable physical abstraction for characterizing the aggregate echo modulation induced by different moving parts of the target, such as UAV paddles, bird wings, vehicle wheels, and pedestrian limbs. 
Even when these scattering points are not spatially resolvable due to limited bandwidth or aperture, their time-varying ranges and radial velocities will still be superimposed in the received echo and manifested as micro-Doppler or time-frequency signatures after coherent processing.}

Specifically,
let us consider that a LAT consists of $L$ scattering points, and 
 ISAC system employs a TRF containing $N$ OFDM symbols to recognize this target.
At the $n$-th OFDM symbol, the Cartesian coordinate position of the $l$-th scattering point in the global coordinate system is represented as $\mathbf{p}_{n,l} = [x_{n,l}, y_{n,l}, z_{n,l}]^T$. Then the {global Cartesian coordinate positions} of all scattering points at the $n$-th OFDM symbol can  constitute a set,  denoted as
\begin{equation}
	\begin{split}
		\begin{aligned}
			\label{deqn_ex1a}
			\mathcal{P}_n = \left\{\mathbf{p}_{n,l} = [x_{n,l}, y_{n,l}, z_{n,l}]^T \mid l=1,2, \dots, L\right\}.
		\end{aligned}
	\end{split}
\end{equation}
During  $N$ OFDM symbols, the LAT undergoes various linear motions  (e.g., near-constant velocity linear motion of the body) and non-linear motions (e.g., rotation of UAV's paddles). These motions cause temporal variations in the spatial positions  of the $L$ scattering points at each OFDM symbol interval, thereby inducing distinct micro-Doppler signatures.
Hence 
we arrange $\mathcal{P}_n$ with $n=0,1,...,N-1$ in an orderly manner over $N$ OFDM symbols to form a \emph{``full-time global Cartesian coordinate positions set"}, denoted as
\begin{equation}
	\begin{split}
		\begin{aligned}
			\label{deqn_ex1a}
			\mathcal{P}_{all} = \left\{\mathcal{P}_0, \mathcal{P}_1,...,\mathcal{P}_{N-1}\right\}.
		\end{aligned}
	\end{split}
\end{equation}

Next,  we establish a local Cartesian coordinate system $O'_k\text{-}x'_k y'_k z'_k$
 for the $k$-th BS, 
in which the origin of the local Cartesian coordinate system is the global Cartesian coordinate position of the $k$-th BS, 
 the positive $x'_k$-axis is aligned with the normal direction of the $k$-th BS's UPA, and the positive $z'_k$-axis is parallel to the positive $z$-axis of the global coordinate system.
The positive   $y'_k$-axis is determined by the right-hand rule relative to the positive $x'_k$-axis and the positive $z'_k$-axis. Associated with the local Cartesian coordinate system of the $k$-th BS, a local spherical coordinate system $O'_k\text{-}r'_k \theta'_k \phi'_k$
 is correspondingly defined to characterize the angular parameters of the target, which satisfies
 \begin{equation}
 	\left\{
 	\begin{aligned}
 		&x'_k = r'_k \cos \phi'_k \cos \theta'_k,\\
 		&y'_k = r'_k \cos \phi'_k \sin \theta'_k,\\
 		&z'_k = r'_k \sin \phi'_k.
 	\end{aligned}
 	\right.
 \end{equation}

Then the global Cartesian coordinate position of the $l$-th scattering point at the $n$-th symbol, i.e., $\mathbf{p}_{n,l} = [x_{n,l}, y_{n,l}, z_{n,l}]^T$, can be transformed into the local spherical coordinate position relative to the $k$-th BS, represented as $\mathbf{p}_{k,n,l}^{\star} = [r_{k,n,l}, \theta_{k,n,l}, \phi_{k,n,l}]^T$, which satisfies
\begin{equation}
	\left\{
	\begin{aligned}
		&r_{k,n,l} = \|\mathbf{p}_{n,l} - \mathbf{p}_k \|_2,\\
		&\theta_{k,n,l}= \frac{\pi}{2}-\arg \left\{
		\frac{\frac{(x_{n,l}-x_k)+\sqrt{-1}\cdot(y_{n,l}-y_k)}{\sqrt{(x_{n,l}-x_k)^2 + (y_{n,l}-y_k)^2}}}
		{\cos \Theta_{k} + \sqrt{-1} \cdot \sin \Theta_{k} } \right\},\\
		&\phi_{k,n,l} =\arcsin \left( 
		\frac{z_{n,l}-z_k}{\|\mathbf{p}_{n,l} - \mathbf{p}_k \|_2}
		 \right),
	\end{aligned}
	\right.
\end{equation}
where  $r_{k,n,l}$, $\theta_{k,n,l}$, and $\phi_{k,n,l}$ indicate the
distance, horizontal angle, pitch angle
 of the $l$-th scattering point relative to the $k$-th BS at the $n$-th OFDM symbol, respectively.
For the $k$-th BS, the {local spherical coordinate positions} of all scattering points of the LAT at the $n$-th OFDM symbol can be aggregated into a set,  denoted as
\begin{equation}
	\begin{split}
		\begin{aligned}
			\label{deqn_ex1a}
			\mathcal{S}_{k,n} = \{ \mathbf{p}_{k,n,l}^{\star} = [r_{k,n,l}, \theta_{k,n,l}, \phi_{k,n,l}]^T | l = 1,2,...,L \}.
		\end{aligned}
	\end{split}
\end{equation}
Then 
we arrange $\mathcal{S}_{k,n}$ with $n=0,1,...,N-1$ in an orderly manner over $N$ OFDM symbols to form a \emph{``full-time local spherical coordinate positions set of the $k$-th BS"}, denoted as
\begin{equation}
	\begin{split}
		\begin{aligned}
			\label{deqn_ex1a}
			\mathcal{S}_{k,all} = \left\{\mathcal{S}_{k,0}, \mathcal{S}_{k,1},...,\mathcal{S}_{k,N-1}\right\}.
		\end{aligned}
	\end{split}
\end{equation}

Based on $\mathcal{S}_{k,n}$ and \cite{11159257}, the sensing echo channel matrix of the LAT on the $m$-th subcarrier of the $n$-th OFDM symbol for the $k$-th BS can be represented as
\begin{equation}
	\begin{split}
		\begin{aligned}
			\label{deqn_ex1a}
			\!\!\!\!\!\!\!\!\!\!\!& \!\!\!\!\! \mathbf{H}_{k,n,m} = \\& \!\!\!\!\!\!  \!\!\!
			\sum_{l=1}^{L} \! \alpha_{k\!,n\!,l}  e^{\!-\!j\! \frac{4\pi\! f_{k\!,\!m}  r_{k\!,\!n\!,\!l}}{c}}\!\mathbf{a}_{\!R,k}(\Psi_{k,n,l}\!,\Omega_{k,n,l})
			\mathbf{a}^T_{\!T,k}(\Psi_{k,n,l}\!,\Omega_{k,n,l}),
		\end{aligned}
	\end{split}
\end{equation}
where $\alpha_{k,n,l}$ represents the channel  fading factor, 
$\Psi_{k,n,l} = 
\cos \phi_{k,n,l}\!\cos \theta_{k,n,l} $ represents the horizontal spatial-domain direction, 
$\Omega_{k,n,l}= \sin \phi_{k,n,l} $ represents the pitch spatial-domain direction, 
$\mathbf{a}_{R,k}(\Psi,\Omega)$ and $\mathbf{a}_{T,k}(\Psi,\Omega)$ are  array  steering vectors for the spatial-domain direction  $(\Psi,\Omega)$ of  the $k$-th BS's RU-UPA and TU-UPA with the form\cite{7523373,9573459}
\begin{align}
	\mathbf{a}_{R,k}(\Psi,\Omega) &=
	\mathbf{a}_{R,k}^{\parallel}(\Psi)\otimes \mathbf{a}_{R,k}^{\perp}(\Omega)
	\in \mathbb{C}^{N_{R,k}\times 1},\\
	\mathbf{a}_{T,k}(\Psi,\Omega) &=
	\mathbf{a}_{T,k}^{\parallel}(\Psi )\otimes \mathbf{a}_{T,k}^{\perp}(\Omega)
	\in \mathbb{C}^{N_{T,k}\times 1}.
\end{align}
Here, $\otimes$ denotes the Kronecker product, and 
\begin{align}
	\!\!\!\!\!\!\!\!\!\mathbf{a}_{R,k}^{\parallel}(\Psi)&\!\!=\!\!\left[1,e^{j\frac{2\pi f_kd\Psi}{c}},...,e^{j\frac{2\pi f_kd\Psi}{c}(N_{R,k}^\parallel-1)}\right]^T \!\!\in\! \mathbb{C}^{N_{R,k}^\parallel\times 1},\\
	\!\!\!\!\!\!\!\!\mathbf{a}_{R,k}^{\perp}(\Omega)&\!\!=\!\! \left[1,e^{j\frac{2\pi f_kd\Omega}{c}},...,e^{j\frac{2\pi f_kd\Omega}{c}(N_{R,k}^\perp-1)}\right]^T \!\! \in\! \mathbb{C}^{N_{R,k}^\perp\times 1},\\
	\!\!\!\!\!\!\!\mathbf{a}_{T,k}^{\parallel}(\Psi)&\!\!=\!\!\left[1,e^{j\frac{2\pi f_kd\Psi}{c}},...,e^{j\frac{2\pi f_kd\Psi}{c}(N_{T,k}^\parallel-1)}\right]^T \!\!\in\! \mathbb{C}^{N_{T,k}^\parallel\times 1},\\
	\!\!\!\!\!\!\!\mathbf{a}_{T,k}^{\perp}(\Omega)&\!=\!\! \left[1,e^{j\frac{2\pi f_kd\Omega}{c}},...,e^{j\frac{2\pi f_kd\Omega}{c}(N_{T,k}^\perp-1)}\right]^T \!\! \in\! \mathbb{C}^{N_{T,k}^\perp\times 1}.
\end{align}

It is worth noting that the echo channel in Eq.~(7) does not explicitly exhibit the velocity variables of the LAT. Instead, the kinematic information, including velocities and positions associated with both linear motions and non-linear motions, is implicitly embedded in the temporal evolution of $\mathcal{P}_{all}$ and  $\mathcal{S}_{k,all}$  across all the $N$ OFDM symbols.

\section{\textcolor{black}{Echo Signals   of Low-Altitude Targets}}

In this section, we formulate the   echo signals for 
UAV, bird, vehicle, and pedestrian
under ISAC network.

\subsection{Motion Process and Echo Channel of the UAV Target}

As shown in Fig.~3, 
 the UAV is modeled as a composite of $L$ scattering points, partitioned into the main body $\mathbb{S}_{U,0}$ and $Q$ rotating paddles $\mathbb{S}_{U,i} (i=1, \dots, Q)$.
 The motion process of the UAV can be  decomposed as follows.
 
 \begin{itemize}
 	
 	\item   1) Paddle rotation update: the $i$-th paddle's scattering points rotate around their respective rotation axes $\mathbf{e}_{U,i}$ with rotation frequency $F_{U,i}$, which can be calculated via Rodrigues' rotation formula\cite{dai2015euler}.

 	\item  2) Attitude update: the spatial attitude of  UAV  is typically characterized by  yaw angle, pitch angle, and roll angle. Then the attitude update process can be calculated by  these attitude angles and 
 	 attitude rotation matrix\cite{9573459}.

 	\item 	3) Position update: the position of the UAV approximately follows uniform linear motion within $N$ OFDM symbols.

 \end{itemize}
More derivation and details of these motion equations for the UAV target   can be found in our  previous work \cite{11159257}.

 By formulating these motion processes, we  obtain the full-time global Cartesian positions set of the UAV target across  $N$ OFDM symbols,  denoted as 
 $\mathcal{P}_{all}^{\rm UAV} = \left\{\mathcal{P}_0^{\rm UAV}, \mathcal{P}_1^{\rm UAV},...,\mathcal{P}_{N-1}^{\rm UAV}\right\}$.  Based on Eq.~(4), we can obtain the full-time local spherical coordinate positions set
   of  the UAV target for 
 the $k$-th BS, denoted as 
 $\mathcal{S}_{k,all}^{\rm UAV} = \left\{\mathcal{S}_{k,0}^{\rm UAV}, \mathcal{S}_{k,1}^{\rm UAV},...,\mathcal{S}_{k,N-1}^{\rm UAV}\right\}$,
\textcolor{black}{where}    
$\mathcal{S}_{k,n}^{\rm UAV} = \{ \mathbf{p}_{k,n,l}^{\star, \rm UAV} = [r_{k,n,l}^{\rm UAV}, \theta_{k,n,l}^{\rm UAV}, \phi_{k,n,l}^{\rm UAV}]^T | l = 1,2,...,L \}.$

\begin{figure}[!t]
	\centering
	\includegraphics[width=90mm]{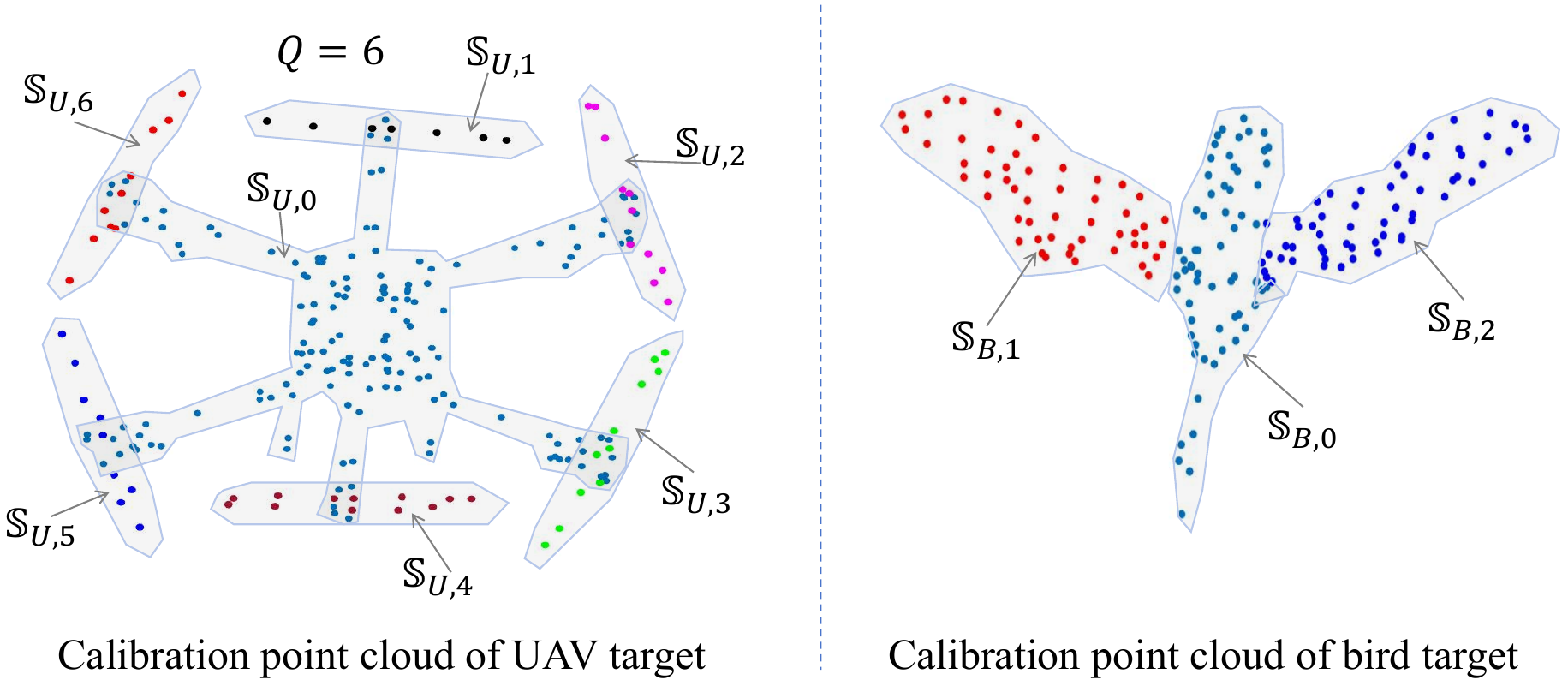}
	\caption{Example of calibration point clouds for UAV target  and  bird target.}
	\label{fig_1}
\end{figure}

Next, 
based on $\mathcal{S}_{k,all}^{\rm UAV}$ and  Eq.~(7),
we can characterize the sensing echo channel matrix of the  UAV target on the $m$-th subcarrier of the $n$-th OFDM symbol for the $k$-th BS   as 
\begin{equation}
	\begin{split}
		\begin{aligned}
			\label{deqn_ex1a}
\!\! \!\!\mathbf{H}_{k,n,m}^{\rm UAV} =  
			\sum_{l=1}^{L}  & \alpha_{k,n,l}^{\rm UAV}  e^{-j \frac{4\pi f_{k,m}  r_{k,n,l}^{\rm UAV}}{c}} \times \\ &
			\mathbf{a}_{R,k}(\Psi_{k,n,l}^{\rm  UAV},\Omega_{k,n,l}^{\rm UAV})
			\mathbf{a}^T_{T,k}(\Psi_{k,n,l}^{\rm UAV},\Omega_{k,n,l}^{\rm UAV}),
		\end{aligned}
	\end{split}
\end{equation}
where $\Psi_{k,n,l}^{\rm UAV} \!=\! \cos \phi_{k,n,l}^{\rm UAV} \cos \theta_{k,n,l}^{\rm UAV}$ and $\Omega_{k,n,l}^{\rm UAV} \!=\! \sin \phi_{k,n,l}^{\rm UAV}$.

\subsection{Motion Process and Echo Channel of the Bird Target}
 
As shown in Fig.~3, 
the bird is modeled as a composite of $L$ scattering points, partitioned into the main body $\mathbb{S}_{B,0}$ and two flapping wings $\mathbb{S}_{B,1}$ and $\mathbb{S}_{B,2}$. 
 The motion process of the bird can be  decomposed as follows.

\begin{itemize}
	
	\item   Wing flapping update: 
	the $i$-th wing's scattering points
	 undergo the  angular-limited rotational motion around their respective rotation axes $\mathbf{e}_{B,i}$. Here the flapping motion is constrained within an angular range $[\beta_{min}, \beta_{max}]$ with the flapping frequency $F_{B,i}$, which can be calculated via Rodrigues' rotation formula\cite{dai2015euler}.

	\item  Attitude update and  position update of the bird   are completely consistent with the UAV.

\end{itemize}
More derivation and details of these motion equations for the bird target   can be found in \cite{11159257}.

\textcolor{black}{Based on these motion process,} we can obtain the full-time global Cartesian positions set of the bird target as 
$\mathcal{P}_{all}^{\rm bird} = \left\{\mathcal{P}_0^{\rm bird}, \mathcal{P}_1^{\rm bird},...,\mathcal{P}_{N-1}^{\rm bird}\right\}$.
The corresponding 
 full-time local spherical coordinate positions set of the bird target for the $k$-th BS can be  denoted as 
$\mathcal{S}_{k,all}^{\rm bird} = \left\{\mathcal{S}_{k,0}^{\rm bird}, \mathcal{S}_{k,1}^{\rm bird},...,\mathcal{S}_{k,N-1}^{\rm bird}\right\}$,
where   
$\mathcal{S}_{k,n}^{\rm bird} = \{ \mathbf{p}_{k,n,l}^{\star, \rm bird} = [r_{k,n,l}^{\rm bird}, \theta_{k,n,l}^{\rm bird}, \phi_{k,n,l}^{\rm bird}]^T | l = 1,2,...,L \}.$
Based on $\mathcal{S}_{k,all}^{\rm bird}$ and 
Eq.~(7), the sensing echo channel matrix of the  bird target on the $m$-th subcarrier of the $n$-th OFDM symbol for the $k$-th BS  can be represented as 
\begin{equation}
	\begin{split}
		\begin{aligned}
			\label{deqn_ex1a}
			\!\! \!\!\mathbf{H}_{k,n,m}^{\rm bird} =  
			\sum_{l=1}^{L}  & \alpha_{k,n,l}^{\rm bird}  e^{-j \frac{4\pi f_{k,m}  r_{k,n,l}^{\rm bird}}{c}} \times \\ &
			\mathbf{a}_{R,k}(\Psi_{k,n,l}^{\rm  bird},\Omega_{k,n,l}^{\rm bird})
			\mathbf{a}^T_{T,k}(\Psi_{k,n,l}^{\rm bird},\Omega_{k,n,l}^{\rm bird}),
		\end{aligned}
	\end{split}
\end{equation}
where $\Psi_{k,n,l}^{\rm bird} \!=\! \cos \phi_{k,n,l}^{\rm bird} \cos \theta_{k,n,l}^{\rm bird}$ and $\Omega_{k,n,l}^{\rm bird} \!=\! \sin \phi_{k,n,l}^{\rm bird}$.

\subsection{Motion Process and Echo Channel of the Vehicle Target}

\begin{figure}[!t]
	\centering
	\includegraphics[width=80mm]{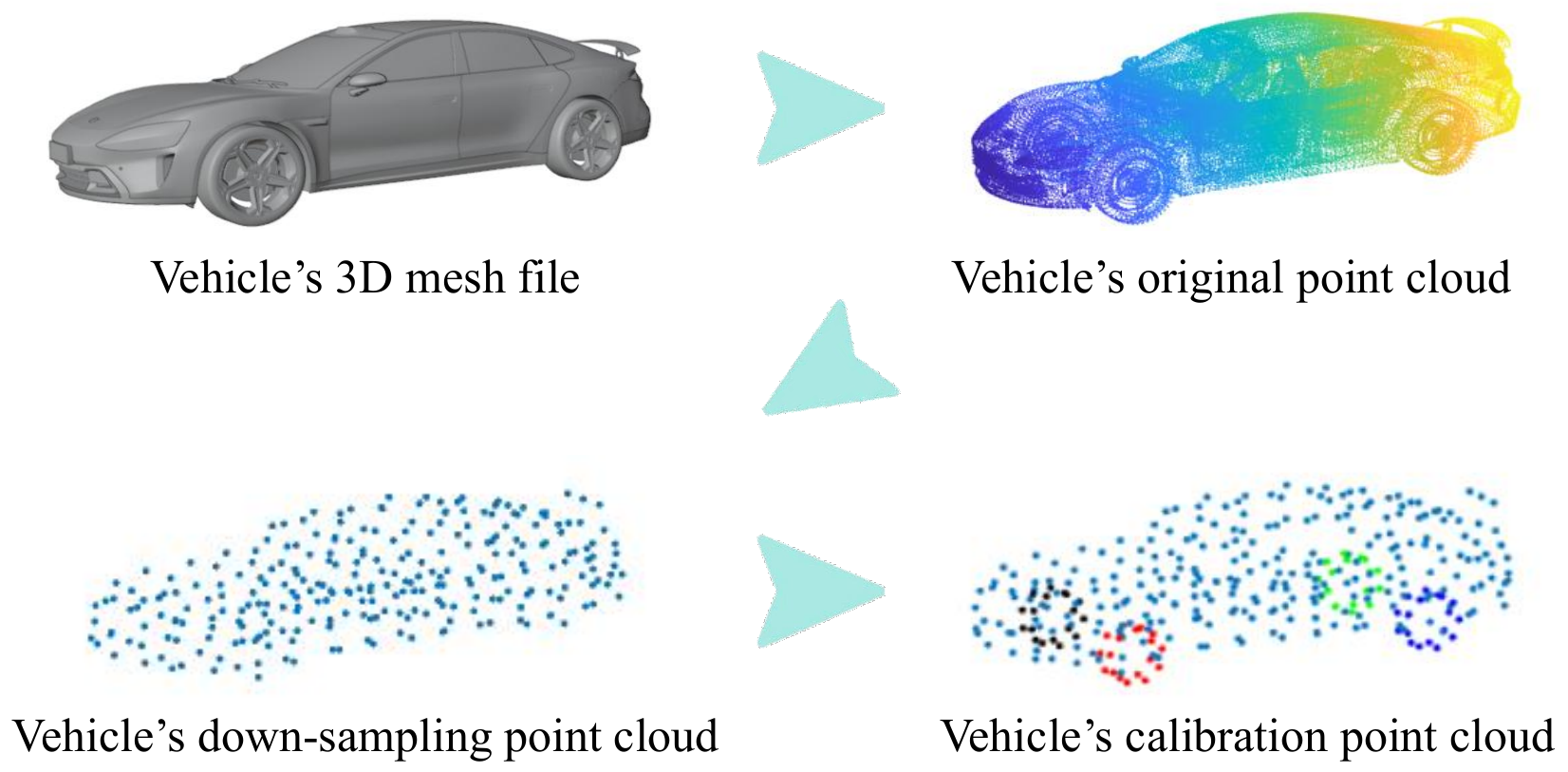}
	\caption{Process diagram to obtain the vehicle's calibration point cloud.}
	\label{fig_1}
\end{figure}

As shown in Fig.~4,
we utilize MeshLab\footnote{\textcolor{black}{MeshLab is an open-source, portable, and extensible software designed for processing and editing large-scale unstructured 3D triangular meshes\cite{mesh1}.}} software to convert the 3D model file of vehicle target into the dense  \emph{original point cloud}.  
Then uniform down-sampling and scaling are applied  to obtain the \emph{down-sampling point cloud} of the vehicle target, which  can remove redundant information while preserving the structural skeleton of the vehicle target.
Since the structures of the wheels are weakened after down-sampling,  \textcolor{black}{it is}  difficult to extract effective wheel data via manual calibration. 
\textcolor{black}{Therefore, based on the actual sizes and relative positions of the wheels, we manually add the point cloud of the wheels to the down-sampling point cloud, thereby obtaining  the
\emph{calibration point cloud (CPC)}
 of the vehicle target,}
denoted as $\mathbb{S}_{V}$, where 
${\rm card} (\mathbb{S}_{V}) = L_V$. 
Let us denote the set of scattering points for the vehicle body as $\mathbb{S}_{V,0}$, and the set of scattering points for each wheel as $\mathbb{S}_{V,1},\mathbb{S}_{V,2},...,\mathbb{S}_{V,W}$, where $W$ represents the total number of wheels on the vehicle, and ${\rm card} \  (\mathbb{S}_{V,i}) = L_{V,i}$, $i = 0,1,...,W$. 
 There are  $\bigcup_{i=0}^{W} \mathbb{S}_{V,i} = \mathbb{S}_V$,  $\mathbb{S}_{V,i} \cap \mathbb{S}_{U,j} = \varnothing$ for $i \neq j \in \{0,1,...,W\}$, and $\sum_{i=0}^W L_{V,i} = L_{V}$.
Meanwhile, denote the position of the $l_{V,i}$-th scattering point in $\mathbb{S}_{V,i}$ as 
$\widetilde{\mathbf{p}}_{l_{V,i},n=-1}$, with $i = 0,1,...,W$  and  $l_{V,i} = 1,...,L_{V,i}$.
Then the CPC of  the vehicle target can be denoted as  $\{\widetilde{\mathbf{p}}_{l_{V,i},n=-1} |  i = 0,1,...,W; l_{V,i} = 1,...,L_{V,i}\}$,
whose center position is roughly located at the origin of the coordinate system.
Specifically, within the vehicle's CPC model, we require the front of the vehicle to be oriented towards the positive $y$-axis.

One of the primary characteristics of vehicle motion is the rotation of the wheels, which can be modeled as the process of scattering points on each wheel rotating around their respective axes. 
Taking the 0-th OFDM symbol as the initial time, we \textcolor{black}{can}  iteratively update the wheel rotation based on the vehicle's CPC model, and then we can obtain 
 the  
 \emph{wheels rotation update point cloud (WRUPC)}   as $\{\widetilde{\mathbf{p}}_{l_{V,i},n} |  i = 0,1,...,W; l_{V,i} = 1,...,L_{V,i}\}$,  $n=0,...,N-1$. 
 Specifically, let $\beta_{V,i,0}$ and $F_{V,i}$ represent the initial phase and rotational frequency of the $i$-th wheel, respectively.
Assume that the rotational frequency of each wheel  is consistent, i.e., $F_{V,1} = F_{V,2} = ... = F_{V,W} = F_{V}$. Meanwhile, 
  there is a direct correspondence between the rotational frequency $F_{V}$ and the vehicle's  forward speed $v_{\rm vehicle}$, which can be expressed as 
\begin{equation}
	\begin{split}
		\begin{aligned}
			\label{deqn_ex1a}
			F_{V} = \frac{v_{\rm vehicle}}{\pi D_{\rm vehicle}},
		\end{aligned}
	\end{split}
\end{equation}
where $D_{\rm vehicle}$ is the diameter of the wheel.
Denote the phase of the $i$-th wheel at the $n$-th   OFDM symbol as $\beta_{V,i,n}$ with 
$i=1,...,W$. We manually designate two points for
the $i$-th wheel as the rotation axis, which are denoted as $\mathbf{p}^a_{V,i}$ and $\mathbf{p}^b_{V,i}$. Then the unit directional vector of the 
rotation axis   is $\mathbf{e}_{V,i} = \frac{\mathbf{p}^a_{V,i} - \mathbf{p}^b_{V,i}}{\left|\mathbf{p}^a_{V,i} - \mathbf{p}^b_{V,i}\right|_2}$. 
When $n \!=\! 0,...,N-1$, 
we calculate the phase change of the $i$-th wheel  as $\Delta \beta_{V,i,n} \!=\! \beta_{V,i,0}$ for $n=0$, and $\Delta \beta_{V,i,n} = 2\pi F_{V} T_r$ for $n=1,...,N-1$.  
\textcolor{black}{Then we  rotate  $\widetilde{\mathbf{p}}_{l_{V,i},n-1}$ around the $i$-th rotation axis by angle $\Delta \beta_{V,i,n}$ to obtain the updated point $\widetilde{\mathbf{p}}_{l_{V,i},n}$.
	According to the Rodrigues' rotation formula\cite{dai2015euler}, 
	$\widetilde{\mathbf{p}}_{l_{V,i},n}$ can be 
	represented as}
\begin{equation}
	\begin{split}
		\begin{aligned}
			\label{deqn_ex1a}
			\widetilde{\mathbf{p}}_{l_{V,i},n} & \!= \mathbf{p}^a_{V,i} + \mathbf{u}_{l_{V,i},n}\cos \Delta \beta_{V,i,n} \\& \quad \! + (\mathbf{e}_{V,i} \times \mathbf{u}_{l_{V,i},n})\sin \Delta \beta_{V,i,n} \\& \quad \!+ \mathbf{e}_{V,i} (\mathbf{e}_{V,i}^T   \mathbf{u}_{l_{V,i},n})(1\!-\! \cos \Delta \beta_{V,i,n}),  
		\end{aligned}
	\end{split}
\end{equation}
where  $\mathbf{u}_{l_{V,i},n} \!=\! \widetilde{\mathbf{p}}_{l_{V,i},n-1} \!-\! \mathbf{p}^a_{V,i}$, $i=1,...,W$.
Meanwhile, the vehicle's body   will not change
in WRUPC, and hence there is $\widetilde{\mathbf{p}}_{l_{V,0},n} = \widetilde{\mathbf{p}}_{l_{V,0},n-1}$.

Next, we need to consider the spatial attitude of the vehicle target, in which the  spatial attitude of one target is typically characterized by yaw angle, pitch angle, and roll angle. 
Assuming that the
 yaw angle $\gamma_{V,Y}$,
 pitch angle $\gamma_{V,P}$, and 
 roll angle  $\gamma_{V,R}$ of  the vehicle target remains unchanged during the observation period.
 At the $n$-th symbol time, 
 we can update  the attitude change from WRUPC to obtain the  vehicle's  \emph{attitude update point cloud} (AUPC)  as $\{\check{\mathbf{p}}_{l_{V,i},n} |  i = 0,1,...,W; l_{V,i} = 1,...,L_{V,i}\}$, which satisfies
\begin{equation}
	\begin{split}
		\begin{aligned}
			\label{deqn_ex1a}
			\check{\mathbf{p}}_{l_{V,i},n} = \mathbf{R}_{Y}(\gamma_{V,Y}) \mathbf{R}_{P}(\gamma_{V,P}) \mathbf{R}_{R}(\gamma_{V,R}) \widetilde{\mathbf{p}}_{l_{V,i},n}
		\end{aligned}
	\end{split}
\end{equation}
with $n=0,...,N-1$.
Here $\mathbf{R}_Y(\delta)  = 
\begin{pmatrix}
	\cos  \delta & -\sin \delta & 0 \\
	\sin \delta  & \cos  \delta & 0 \\
	0 & 0  & 1
\end{pmatrix}$,
$\mathbf{R}_P(\delta ) \!=\! 
\begin{pmatrix}
	1 & 0 & 0 \\
	0 & \cos  \delta  & -\sin \delta   \\
	0 & \sin \delta   & \cos  \delta  
\end{pmatrix}$,
$\mathbf{R}_R(\delta) \!=\! 
\begin{pmatrix}
	\cos  \delta & 0 &  \sin \delta   \\
	0 & 1  & 0\\
	- \sin \delta  & 0 & \cos \delta
\end{pmatrix}$
are  attitude rotation matrices\footnote{In this work, the front orientation of the vehicle target  in the  CPC model is $[0,1,0]^T$. Here $\gamma_{V,Y}$, $\gamma_{V,P}$, $\gamma_{V,R}$,
$\mathbf{R}_Y(\delta)$, $\mathbf{R}_P(\delta)$,
and $\mathbf{R}_R(\delta )$ are 
	 all defined based on the front orientation of the vehicle target.}\cite{9573459}.

Subsequently, we can update  the position change from AUPC to obtain the  vehicle's  \emph{position update point cloud} (PUPC)  as $\{{\mathbf{p}}_{l_{V,i},n} |  i = 0,1,...,W; l_{V,i} = 1,...,L_{V,i}\}$. 
Note that the velocity vector of the vehicle should be parallel to the direction of the vehicle's front.
Hence the velocity vector of the vehicle should be 
\begin{equation}
	\begin{split}
		\begin{aligned}
			\label{deqn_ex1a}
			\!\!\!\!\!\mathbf{v}_{\rm vehicle} =
			v_{\rm vehicle} \cdot \mathbf{R}_{Y}(\gamma_{V,Y}) \mathbf{R}_{P}(\gamma_{V,P}) \mathbf{R}_{R}(\gamma_{V,R}) \begin{pmatrix}
				0  \\
				1  \\
				0  
			\end{pmatrix}.
		\end{aligned}
	\end{split}
\end{equation} 
 When $n = 0,...,N-1$,  calculate the position of the vehicle at the $n$-th symbol  as $\mathbf{p}_{{\rm vehicle},n} =\mathbf{p}_{{\rm vehicle},0} +  nT_r  \mathbf{v}_{\rm vehicle}$,
 where $\mathbf{p}_{{\rm vehicle},0}$ is the initial position. 
  Then 
the  point $\mathbf{p}_{l_{V,i},n}$
is 
\begin{equation}
	\begin{split}
		\begin{aligned}
			\label{deqn_ex1a}
			\mathbf{p}_{l_{V,i},n} = \check{\mathbf{p}}_{l_{V,i},n} + \mathbf{p}_{{\rm vehicle},n}.
		\end{aligned}
	\end{split}
\end{equation}
Here, $\{{\mathbf{p}}_{l_{V,i},n} |  i = 0,1,...,W; l_{V,i} = 1,...,L_{V,i}\} = \mathcal{P}_n^{\rm vehicle}$ is precisely
the {global Cartesian coordinate positions set} of the vehicle target. 

Eventually, we obtain the full-time global Cartesian positions set of the vehicle target across  $N$ OFDM symbols, i.e., 
$\mathcal{P}_{all}^{\rm vehicle} = \left\{\mathcal{P}_0^{\rm vehicle}, \mathcal{P}_1^{\rm vehicle},...,\mathcal{P}_{N-1}^{\rm vehicle}\right\}$.  Based on Eq.~(4), we can obtain the full-time local spherical coordinate positions set of the vehicle target for the $k$-th BS,  denoted as 
$\mathcal{S}_{k,all}^{\rm vehicle} = \left\{\mathcal{S}_{k,0}^{\rm vehicle}, \mathcal{S}_{k,1}^{\rm vehicle},...,\mathcal{S}_{k,N-1}^{\rm vehicle}\right\}$, where   
$\mathcal{S}_{k,n}^{\rm vehicle} = \{ \mathbf{p}_{k,n,l}^{\star, \rm vehicle} = [r_{k,n,l}^{\rm vehicle}, \theta_{k,n,l}^{\rm vehicle}, \phi_{k,n,l}^{\rm vehicle}]^T | l = 1,2,...,L \}.$

Next, 
based on $\mathcal{S}_{k,all}^{\rm vehicle}$ and  Eq.~(7),
we can characterize the sensing echo channel matrix of the  vehicle target on the $m$-th subcarrier of the $n$-th OFDM symbol for the $k$-th BS   as 
\begin{equation}
	\begin{split}
		\begin{aligned}
			\label{deqn_ex1a}
			\!\!\!\!\!\! \!\!\!\!\mathbf{H}_{k,n,m}^{\rm vehicle} \!\!=  \!\!
			\sum_{l=1}^{L}  & \alpha_{k,n,l}^{\rm vehicle}  e^{-j \frac{4\pi f_{k,m}  r_{k,n,l}^{\rm vehicle}}{c}} \times \\ &
			\!\!\!\!\!\!\!\!
			\mathbf{a}_{R,k}(\Psi_{k,n,l}^{\rm  vehicle},\Omega_{k,n,l}^{\rm vehicle})
			\mathbf{a}^T_{T,k}(\Psi_{k,n,l}^{\rm vehicle},\Omega_{k,n,l}^{\rm vehicle}),
		\end{aligned}
	\end{split}
\end{equation}
where $\Psi_{k,n,l}^{\rm vehicle} \!\!\!=\!\!\! \cos \! \phi_{k,n,l}^{\rm vehicle}\! \cos \! \theta_{k,n,l}^{\rm vehicle}$ and $\Omega_{k,n,l}^{\rm vehicle} \!\!=\!\! \sin \! \phi_{k,n,l}^{\rm vehicle}$.

\begin{figure}[!t]
	\centering
	\includegraphics[width=88mm]{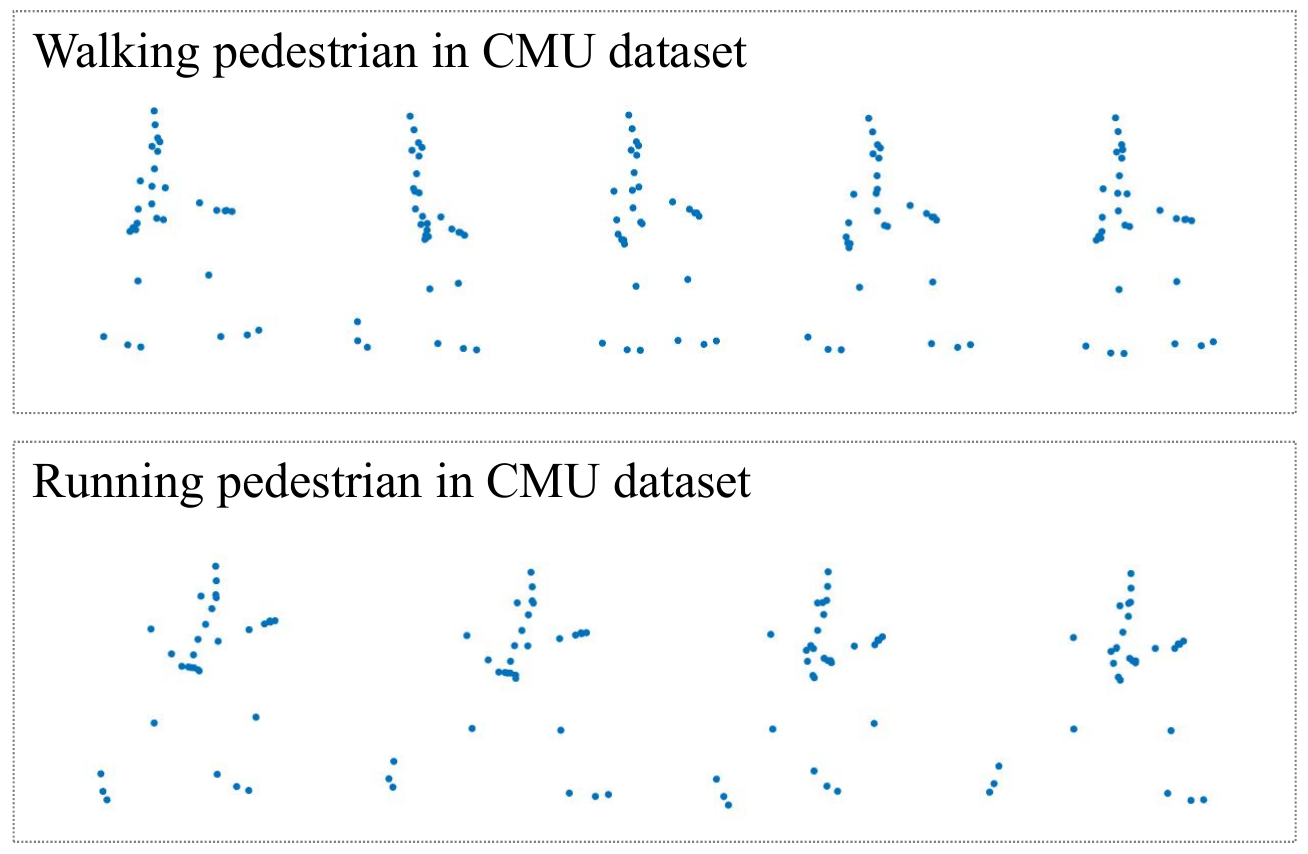}
	\caption{Examples of pedestrian point cloud in CMU dataset.}
	\label{fig_1}
\end{figure}

\subsection{Motion Process and Echo Channel of the Pedestrian Target}

As shown in Fig.~5, 
we employ the classic public human motion dataset from  Carnegie Mellon University (CMU)
  as the  data support for the pedestrian motion process.  The CMU dataset \cite{CMUmocap} is collected in the  controlled environment using 12 Vicon MX-40 infrared cameras (4.0 MP) operating at 120 Hz within the $3\text{ m} \times 8\text{ m}$ area. By tracking the markers worn on the human and applying spatial geometric algorithms, CMU extracts the 3D coordinates and rotation data for 31 key nodes (e.g., head, chest,  limbs, etc.) of one  human target across the  motion sequence.
  Here we employ these 31 key nodes of one human as the CPC of the pedestrian target, denoted as 
  $\mathbb{S}_{P}$, where 
  ${\rm card} (\mathbb{S}_{P}) = L_P$.
The position of the $l_P$-th scattering point in $\mathbb{S}_{P}$ is denoted as $\widetilde{\mathbf{p}}_{l_{P},n=-1}$, where $l_{P} = 1,...,L_{P}$. 
Then the CPC of the pedestrian target 
 can be represented as $\{\widetilde{\mathbf{p}}_{l_{P},n=-1} |   l_{P} = 1,...,L_{P}\}$, 
whose center position is roughly located at the origin of the coordinate system.
Specifically, within the pedestrian's CPC model, we also require the front of the pedestrian to be oriented towards the positive $y$-axis.

Next,
we construct the motion process of pedestrian target based on the CMU dataset.
Taking the $0$-th OFDM symbol as the initial time, we address the sampling interval mismatch between the $120\text{ Hz}$ CMU data and the OFDM symbols in this work via linear interpolation. 
Specifically, in CMU dataset, 
let $T_{\rm CMU}$ denote 
the time interval between adjacent original sampling frames,  let $t_h = h \cdot T_{\rm CMU}$ denote the physical time corresponding to the $h$-th original sampling frame, and let $\widetilde{\mathbf{p}}_{l_{P},h}^{\rm CMU}$ denote the  position of the $l_P$-th scattering point of the human in the $h$-th   original sampling frame. 
\textcolor{black}{Meanwhile,}
the physical time corresponding to the $n$-th OFDM symbol is $t_n = n \cdot T_{r}$.
When $t_h \leq t_n < t_{h+1}$, the position of the $l_{P}$-th scattering point within the pedestrian target 
 at the $n$-th OFDM symbol  can be represented as 
\begin{equation}
	\begin{split}
		\begin{aligned}
			\label{deqn_ex1a}
\widetilde{\mathbf{p}}_{l_{P},n} =  \widetilde{\mathbf{p}}_{l_{P},h}^{\rm CMU} + \frac{t_n-t_h}{t_{h+1}-t_h}\cdot \left( \widetilde{\mathbf{p}}_{l_{P},{h+1}}^{\rm CMU} -
\widetilde{\mathbf{p}}_{l_{P},h}^{\rm CMU} \right).
		\end{aligned}
	\end{split}
\end{equation}
Hence 
we can obtain the \emph{limb motion point cloud (LMPC)}
 of the pedestrian target  at the $n$-th OFDM symbol
 as  $\{\widetilde{\mathbf{p}}_{l_{P},n} |   l_{P} = 1,...,L_{P}\}$ with $n=0,1,...,N-1$.

Subsequently, we  
 characterize the attitude update and position update of  pedestrian target based on  LMPC model.
Assume that the
yaw angle $\gamma_{P,Y}$,
pitch angle $\gamma_{P,P}$, and 
roll angle  $\gamma_{P,R}$ of  pedestrian target remains unchanged during the observation period. Then the pedestrian's  AUPC can be represented   as $\{\check{\mathbf{p}}_{l_{P},n} |  l_{P} = 1,...,L_{P}\}$, which satisfies
$\check{\mathbf{p}}_{l_{P},n} = \mathbf{R}_{Y}(\gamma_{P,Y}) \mathbf{R}_{P}(\gamma_{P,P}) \mathbf{R}_{R}(\gamma_{P,R}) \widetilde{\mathbf{p}}_{l_{P},n}$.
Assume that the initial position and forward speed of the pedestrian  are $v_{\rm pedestrian}$ and $\mathbf{p}_{{\rm pedestrian},0}$. 
Note that the velocity vector of  pedestrian should be parallel to the direction of  pedestrian's front.
The velocity vector of  pedestrian should be 
$\mathbf{v}_{\rm pedestrian} =
v_{\rm pedestrian}  \mathbf{R}_{Y}(\gamma_{P,Y}) \mathbf{R}_{P}(\gamma_{P,P}) \mathbf{R}_{R}(\gamma_{P,R})
(0,1,0)^T$. When $n$ $ = 0,...,N-1$, we  can  calculate the position of the pedestrian at the $n$-th symbol  as $\mathbf{p}_{{\rm pedestrian},n} =\mathbf{p}_{{\rm pedestrian},0} +  nT_r  \mathbf{v}_{\rm pedestrian}$. 
Then the  pedestrian's  \emph{position update point cloud} (PUPC) can be obtained  as $\{{\mathbf{p}}_{l_{P},n} |  l_{P} = 1,...,L_{P}\}$,
which satisfies 
$\mathbf{p}_{l_{P},n} = \check{\mathbf{p}}_{l_{P},n} + \mathbf{p}_{{\rm pedestrian},n}.$
Here  the PUPC is precisely
the {global Cartesian coordinate positions set} of the pedestrian target, i.e., 
$\mathcal{P}_n^{\rm pedestrian} = \{{\mathbf{p}}_{l_{P},n} |   l_{P} = 1,...,L_{P}\}$. 

Then we  obtain the full-time global Cartesian positions set of the pedestrian target across  $N$ OFDM symbols as 
$\mathcal{P}_{all}^{\rm pedestrian} = \left\{\mathcal{P}_0^{\rm pedestrian}, \mathcal{P}_1^{\rm pedestrian},...,\mathcal{P}_{N-1}^{\rm pedestrian}\right\}$.
Based on Eq.~(4), we can obtain the full-time local spherical coordinate positions set of the pedestrian target for the $k$-th BS as 
$\mathcal{S}_{k,all}^{\rm pedestrian} = \left\{\mathcal{S}_{k,0}^{\rm pedestrian}, \mathcal{S}_{k,1}^{\rm pedestrian},...,\mathcal{S}_{k,N-1}^{\rm pedestrian}\right\}$, where  
$\mathcal{S}_{k,n}^{\rm pedestrian} = \{ \mathbf{p}_{k,n,l}^{\star, \rm pedestrian} = [r_{k,n,l}^{\rm pedestrian}, \theta_{k,n,l}^{\rm pedestrian}, \phi_{k,n,l}^{\rm pedestrian}]^T | l = 1,2,...,L \}.$
Then we can characterize the sensing echo channel matrix of the  pedestrian target on the $m$-th subcarrier of the $n$-th OFDM symbol for the $k$-th BS   as 
\begin{equation}
	\begin{split}
		\begin{aligned}
			\label{deqn_ex1a}
			\!\!\!\!\! \!\!\mathbf{H}_{k,n,m}^{\rm pedestrian} =  
			\sum_{l=1}^{L}  & \alpha_{k,n,l}^{\rm pedestrian}  e^{-j \frac{4\pi f_{k,m}  r_{k,n,l}^{\rm pedestrian}}{c}} \times \\ & \!\!\!\!\!\!\!\!\!\!\!\!\!\!\!
			\!\!\!\!\!\!\!\!\!\!\!\!\!\!\!
			\!\!\!\!\!\!\!\!\!\!\!\!\!\!\!
			\mathbf{a}_{R,k}(\Psi_{k,n,l}^{\rm  pedestrian},\Omega_{k,n,l}^{\rm pedestrian})
			\mathbf{a}^T_{T,k}(\Psi_{k,n,l}^{\rm pedestrian},\Omega_{k,n,l}^{\rm pedestrian}),
		\end{aligned}
	\end{split}
\end{equation}
where $\Psi_{k,n,l}^{\rm pedestrian} \!=\! \cos \phi_{k,n,l}^{\rm pedestrian} \cos \theta_{k,n,l}^{\rm pedestrian}$ and $\Omega_{k,n,l}^{\rm pedestrian} \!=\! \sin \phi_{k,n,l}^{\rm pedestrian}$.

\subsection{Echo Signals of  Low-Altitude Targets}

The transmitted signal on the $m$-th subcarrier of the $n$-th OFDM symbol from the $k$-th BS can be represented as ${\mathbf{x}}_{k,n,m} =  \sqrt{{\rho}_{s,k}P_{T,k}/N_{T,k}}\mathbf{a}_{T,k}\left(\Psi_{T,k},\Omega_{T,k}\right)s_{k,n,m}$,
where  $P_{T,k}$ and ${\rho}_{s,k}$ are the transmission power and the power allocation factor towards the sensing direction 
 of the $k$-th BS, respectively,
 $\left(\Psi_{T,k},\Omega_{T,k}\right)$ is the beamforming direction, and  $s_{k,n,m}$ is the sensing detection signal. 
Meanwhile, the   $k$-th BS also performs receive beamforming with $\mathbf{w}_{k,n,m} = \sqrt{{1}/{N_{R,k}}} \mathbf{a}_{R,k}\left(\Psi_{R,k},\Omega_{R,k}\right)$. 
Then the frequency-domain echo signal received by  the $k$-th BS on the $m$-th subcarrier of the $n$-th symbol can be represented as\footnote{\textcolor{black}{In practical sub-6 GHz ISAC networks, the received echo may also be affected by  multi-path propagation, hardware impairments, synchronization errors, and imperfect range-bin selection. 
This paper focuses on the target-dependent echo components and the recognition gains brought by multi-scale feature extraction and multi-BS collaborative observations. 
The above non-ideal propagation and system effects can be incorporated into the received signal as additional interference or perturbation terms, and their comprehensive modeling is left for future extensions.}}
\begin{equation}
	\begin{split}
		\begin{aligned}
			\label{deqn_ex1a}
y_{k,n,m} = \mathbf{w}_{k,n,m}^H \mathbf{H}_{k,n,m}^{\rm target} {\mathbf{x}}_{k,n,m}^* + n_{k,n,m},
		\end{aligned}
	\end{split}
\end{equation}
where $n_{k,n,m}$ is  zero-mean additive   Gaussian   noise  with variance  $\sigma^2$.
Moreover, 
$\mathbf{H}_{k,n,m}^{\rm target} = \mathbf{H}_{k,n,m}^{\rm UAV}$ when the target is a UAV,  $\mathbf{H}_{k,n,m}^{\rm target} = \mathbf{H}_{k,n,m}^{\rm bird}$ when the target is a bird,
$\mathbf{H}_{k,n,m}^{\rm target} = \mathbf{H}_{k,n,m}^{\rm vehicle}$ when the target is a vehicle,
and $\mathbf{H}_{k,n,m}^{\rm target} = \mathbf{H}_{k,n,m}^{\rm pedestrian}$ when the target is a pedestrian.

We command the transmitting  beam direction and receiving beam direction of the $k$-th BS to directly point towards the initial direction of the LAT. After receiving the echo signal $y_{k,n,m}$, we  erase the influence of $s_{k,n,m}$ and obtain
$\check{y}_{k,n,m} = y_{k,n,m}/s_{k,n,m}$.
Then  we concatenate the  echo signals on  all subcarriers and  all OFDM symbols for the $k$-th BS  into a matrix $\mathbf{Y}_k \in \mathbb{C}^{N\times M}$, where $\mathbf{Y}_k[n,m] = \check{y}_{k,n,m}$, $n=0,...,N-1$, and $k \in \{ A,B,C\}$.

\section{
	Multi-BS and Multi-Scale Feature Fusion Enabled 
	Low-Altitude Target Recognition Scheme}

In this section, we propose the
multi-BS and multi-scale feature fusion enabled 
LAT recognition scheme.

\begin{figure*}[!t]
	\centering
	\subfloat[]{\includegraphics[width=40mm,height=33mm]{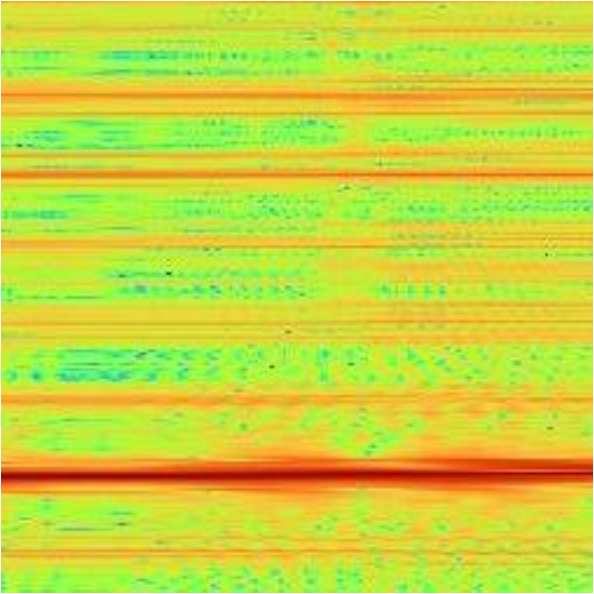}%
		\label{fig_first_case}}
	\hfil
	\subfloat[]{\includegraphics[width=40mm,height=33mm]{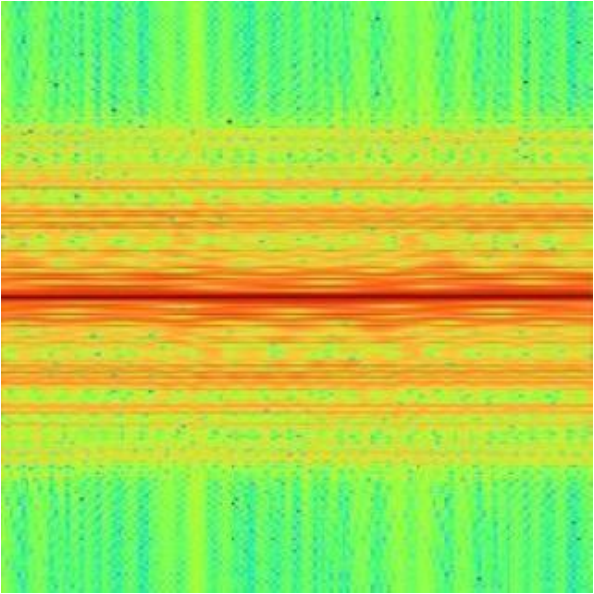}%
		\label{fig_first_case}}
	\hfil
	\subfloat[]{\includegraphics[width=40mm,height=33mm]{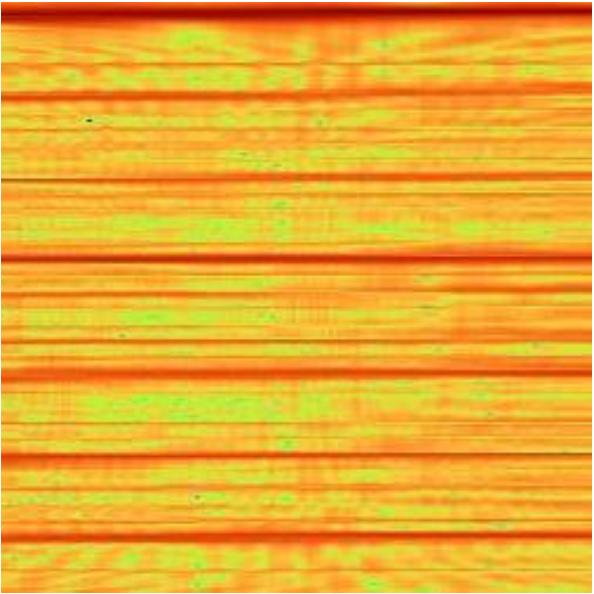}%
		\label{fig_first_case}}
	\hfil
	\subfloat[]{\includegraphics[width=40mm,height=33mm]{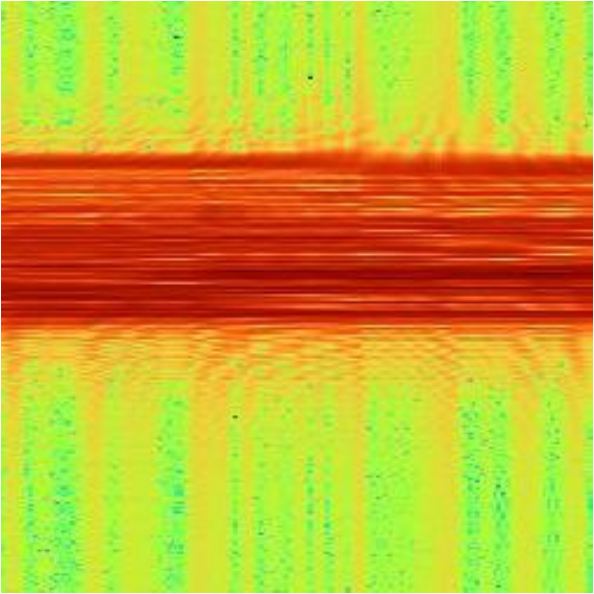}%
		\label{fig_second_case}}
	\caption{\textcolor{black}{
			VRP-TF spectrum observed by BS-A for four categories of LATs, in which the horizontal axis represents the time, and the vertical axis represents the Doppler units.
			(a) UAV.
			(b) Bird.
			(c) Vehicle.
			(d) Pedestrian.}}
	\label{fig_sim}
\end{figure*}

\subsection{Symbol Domain Complex Signal of Low-Altitude Target}

Let us perform $M$-point inverse discrete Fourier transform (IDFT)   on $\mathbf{Y}_k \in \mathbb{C}^{N\times M}$ in the distance dimension, i.e.,  perform $M$-points IDFT   on each row of $\mathbf{Y}_k \in \mathbb{C}^{N\times M}$ to obtain $\mathbf{Y}_{k,r} \in \mathbb{C}^{N\times M}$  \textcolor{black}{with}
\begin{equation}
	\begin{split}
		\begin{aligned}
			\label{deqn_ex1a}
			\mathbf{Y}_{k,r}[n,m_0] =  \frac{1}{M}\sum_{i_0 = 0}^{M-1}\mathbf{Y}_{k}[n,i_0]e^{j\frac{2\pi i_0}{M}m_0},
		\end{aligned}
	\end{split}
\end{equation}
where $m_0 = 0,1,...,M-1$ and $n=0,1,...,N-1$.
Then we take the modulo value of each element of $\mathbf{Y}_{k,r} \in \mathbb{C}^{N\times M}$ to obtain $\mathbf{Y}_{k,r,\rm abs} \in \mathbb{C}^{N\times M}$, which satisfies
$\mathbf{Y}_{k,r,\rm abs}[n,m] = \left| \mathbf{Y}_{k,r}[n,m] \right|$, $m = 0,1,...,M-1$ and $n=0,1,...,N-1$. 
For the $k$-th BS, the index number of the distance unit where the LAT is located at the $n$-th OFDM symbol  can be represented as
$m_{k,n,{\rm target}} = \mathop{\arg\max}\limits_{m} \mathbf{Y}_{k,r,\rm abs}[n,:]$. Then we can extract the symbol domain complex signal vector of the LAT from $\mathbf{Y}_{k,r} \in \mathbb{C}^{N\times M}$ as $\mathbf{Y}_{k,{\rm bin}} \in \mathbb{C}^{N\times 1}$ with 
$\mathbf{Y}_{k,{\rm bin}} [n] = \mathbf{Y}_{k,r} [n,m_{k,n,{\rm target}}]$. 
Subsequently, we only need to perform signal processing on $\mathbf{Y}_{k,{\rm bin}}$ to recognize the  target.

\subsection{\textcolor{black}{Perform STFT on Symbol Domain Complex Signal}}

Short-Time Fourier Transform (STFT) is one of the fundamental methods in time-frequency analysis\cite{gabor1946theory,1455039}. The basic steps \textcolor{black}{to perform} STFT on $\mathbf{Y}_{k,{\rm bin}} \in \mathbb{C}^{N\times 1}$ are as follows.

Let us set the window length of STFT as $L_{\rm win}$, set the sliding step size as $\Delta L_{\rm win}$, and set the number of FFT points within each window as $P_{\rm win}$. 
Then we can divide $\mathbf{Y}_{k,{\rm bin}} \in \mathbb{C}^{N\times 1}$ into $G = \lfloor \frac{N-L_{\rm win}}{\Delta L_{\rm win}} \rfloor$ groups, which consists of ordered and overlapping signal segments. The $g$-th group of signal segments is $\mathbf{Y}_{k,{g}} = 
\mathbf{Y}_{k,{\rm bin}}
[(g-1)\Delta L_{\rm win}+1:
(g-1)\Delta L_{\rm win} + L_{\rm win}]
 \in \mathbb{C}^{L_{\rm win} \times 1}$.
 Next, we perform \textcolor{black}{$P_{\rm win}$-point} FFT  on $\mathbf{Y}_{k,{g}}$, and move the zero-frequency component to the center position of the result vector, thereby obtaining $\mathbf{Y}_{k,g,\rm FFT}
  \in \mathbb{C}^{P_{\rm win} \times 1}$. 
Take the modulo value of each element of $\mathbf{Y}_{k,g,\rm FFT}$ to obtain $\mathbf{Y}_{k,g,\rm FFT,abs}  \in \mathbb{R}^{P_{\rm win} \times 1}$, which satisfies $\mathbf{Y}_{k,g,\rm  FFT,abs}[p] = \left| \mathbf{Y}_{k,g,\rm FFT}[p] \right|$ with $p=1,...,P_{\rm win}$. 
Then we can concatenate $\mathbf{Y}_{k,1,\rm FFT,abs}, \mathbf{Y}_{k,2,\rm FFT,abs},...,\mathbf{Y}_{k,G,\rm FFT,abs}$ into a matrix as 
\begin{equation}
	\begin{split}
		\begin{aligned}
			\label{deqn_ex1a}
			\mathbf{F}_{k,\rm STFT} = \left[ 
			\mathbf{Y}_{k,1,\rm FFT,abs}, ...,\mathbf{Y}_{k,G,\rm FFT,abs}
			\right] \in \mathbb{R}^{P_{\rm win} \times G},
		\end{aligned}
	\end{split}
\end{equation}
which \textcolor{black}{is exactly}  the result of performing STFT on $\mathbf{Y}_{k,{\rm bin}}$. 
\textcolor{black}{The above steps of STFT can be summarized  as}
\begin{equation}
	\begin{split}
		\begin{aligned}
			\label{deqn_ex1a}
			\mathbf{F}_{k,\rm STFT} = {\rm STFT} \left\{
			\mathbf{Y}_{k,{\rm bin}},
			L_{\rm win}, \Delta L_{\rm win}, P_{\rm win}
			 \right\}.
		\end{aligned}
	\end{split}
\end{equation}

Although the STFT-based time-frequency spectrum of LAT can be obtained by Eq.~(27), selecting appropriate pre-parameters $L_{\rm win}, \Delta L_{\rm win}, P_{\rm win}$ requires comprehensive consideration of various factors such as the time-frequency resource allocation of the BS, the typical micro-motion frequency of LAT, the velocity resolution requirements of time-frequency analysis, and the time resolution requirements of time-frequency analysis.

Generally, STFT faces conflicts and trade-offs between velocity resolution and time resolution. 
On the one hand,  if $L_{\rm win}$ is set to a large value and $P_{\rm win} = L_{\rm win}$, \textcolor{black}{then}  we can attain a high velocity resolution within one single window, thereby enabling a better observation for the velocity details of LAT.
However, a longer observation window may far exceed a single micro-motion period of the target. This tends to smear the periodic patterns of the micro-motion, making them less discernible and resulting in a significant loss of time resolution. 
On the other hand, if $L_{\rm win}$ is set to a small value and $P_{\rm win} = L_{\rm win}$, 
the observation period of one window will be  smaller than the micro-motion period of the LAT. 
Then 
 obvious micro-motion periodicity of LAT can be observed in the STFT spectrum, indicating a high time resolution. 
 However, due to the small value of $L_{\rm win}$, the velocity resolution within one single window is very low, resulting in the loss of detail information about target velocity.

 \begin{figure*}[!t]
 	\centering
 	\subfloat[]{\includegraphics[width=40mm,height=33mm]{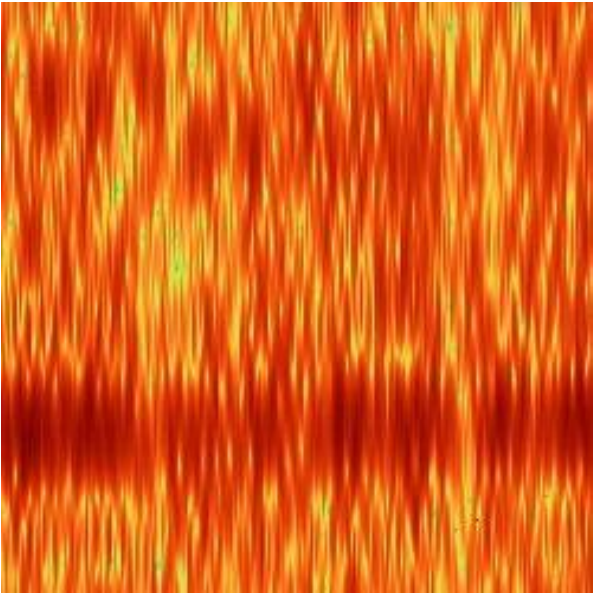}%
 		\label{fig_first_case}}
 	\hfil
 	\subfloat[]{\includegraphics[width=40mm,height=33mm]{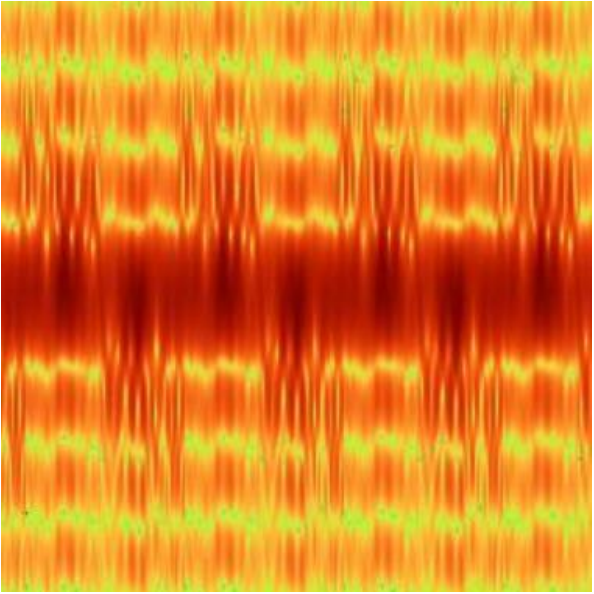}%
 		\label{fig_first_case}}
 	\hfil
 	\subfloat[]{\includegraphics[width=40mm,height=33mm]{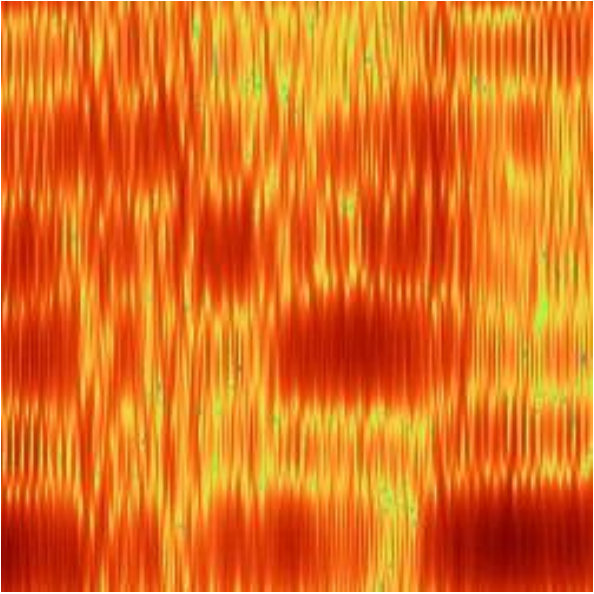}%
 		\label{fig_first_case}}
 	\hfil
 	\subfloat[]{\includegraphics[width=40mm,height=33mm]{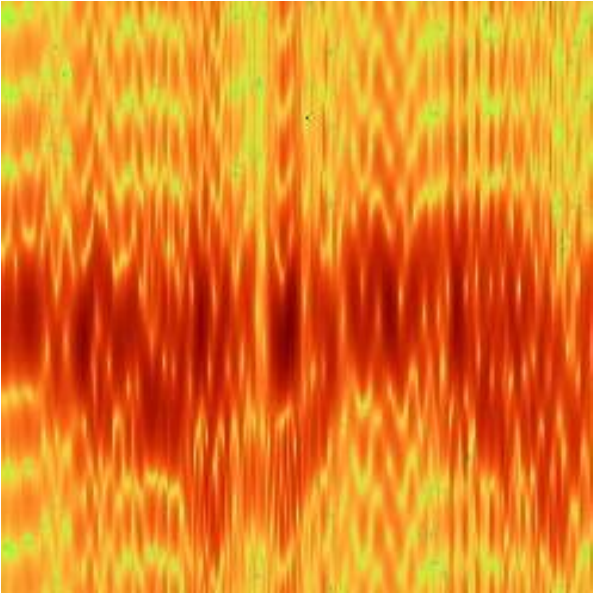}%
 		\label{fig_second_case}}
 	\caption{TRP-TF spectrum observed by BS-A for four categories of LATs, in which the horizontal axis represents the time, and the vertical axis represents the Doppler units.
 		(a) UAV.
 		(b) Bird.
 		(c) Vehicle.
 		(d) Pedestrian.}
 	\label{fig_sim}
 \end{figure*}

 \begin{figure*}[!t]
 	\centering
 	\subfloat[]{\includegraphics[width=40mm,height=33mm]{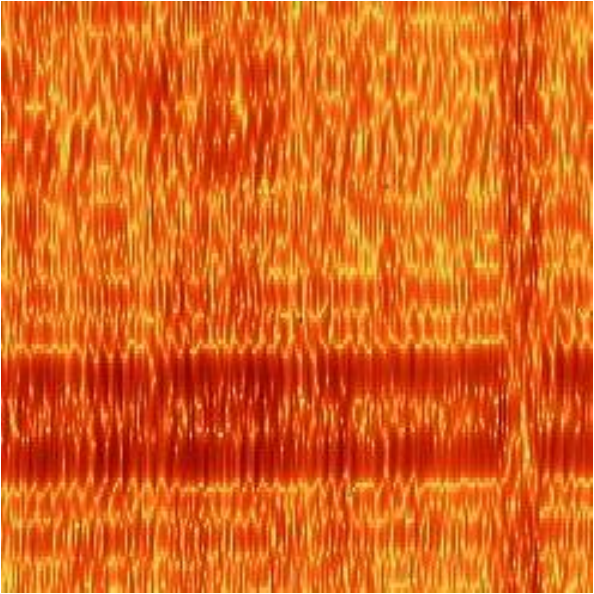}%
 		\label{fig_first_case}}
 	\hfil
 	\subfloat[]{\includegraphics[width=40mm,height=33mm]{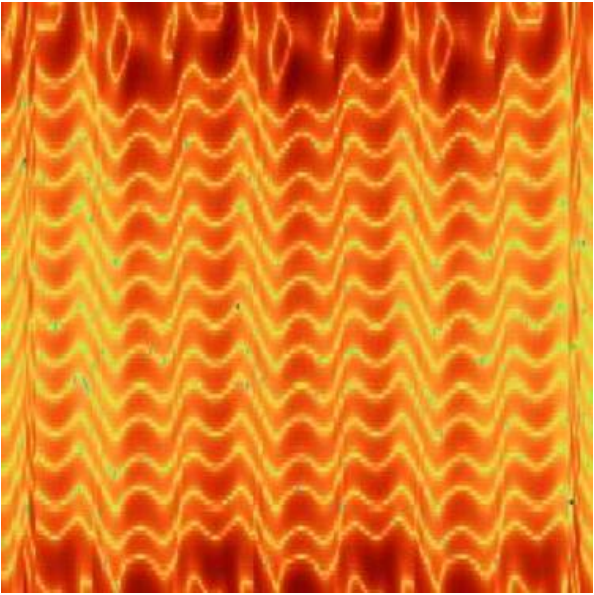}%
 		\label{fig_first_case}}
 	\hfil
 	\subfloat[]{\includegraphics[width=40mm,height=33mm]{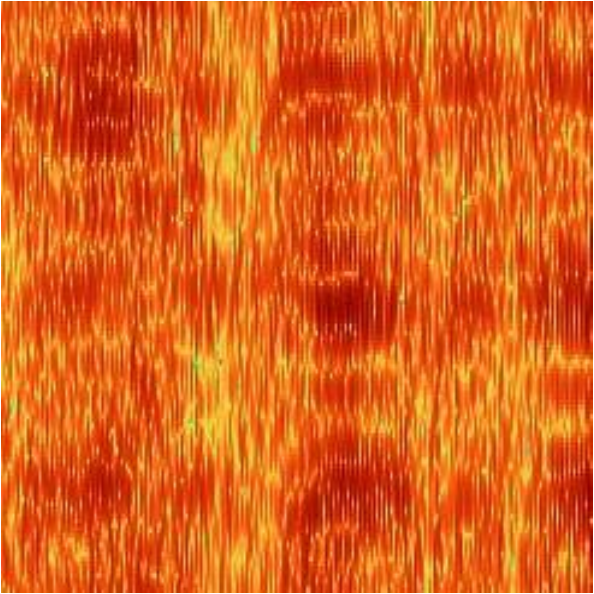}%
 		\label{fig_first_case}}
 	\hfil
 	\subfloat[]{\includegraphics[width=40mm,height=33mm]{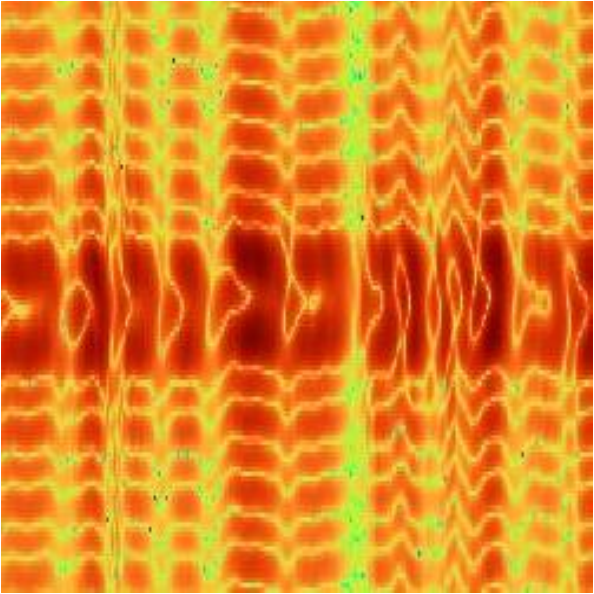}%
 		\label{fig_second_case}}
 	\caption{VT-TF spectrum observed by BS-A for four categories of LATs, in which the horizontal axis represents the time, and the vertical axis represents the velocity matching filtering units.
 		(a) UAV.
 		(b) Bird.
 		(c) Vehicle.
 		(d) Pedestrian.}
 	\label{fig_sim}
 \end{figure*}
 
 \subsection{Velocity-Resolution-Preferred Time-Frequency Spectrum}

   Let us set $L_{\rm win}$ to a large value $L_{\rm win,long}$, set $\Delta L_{\rm win} = \Delta L_{\rm win,long}$, and set $ P_{\rm win} =  P_{\rm win,long}$. Then we can obtain
   the \emph{Velocity-Resolution-Preferred Time-Frequency Spectrum (VRP-TF spectrum)} of the  LAT from the 
   $k$-th BS as
   \begin{equation}
   	\begin{split}
   		\begin{aligned}
   			\label{deqn_ex1a}
   			\!\!\!\!
   			\mathbf{F}_{k,\rm STFT}^{\rm VRP} \!=\! {\rm STFT} \left\{
   			\mathbf{Y}_{k,{\rm bin}},
   			L_{\rm win,long}, \Delta L_{\rm win,long}, P_{\rm win,long}
   			\right\}.
   		\end{aligned}
   	\end{split}
   \end{equation}

Fig.~6  shows the examples of the VRP-TF spectrum observed by BS-A for four categories of LATs. 
The VRP-TF spectrum possesses  high velocity resolution. It can be seen that there are obvious intermittent horizontal stripes in the VRP-TF spectrum of UAV and vehicle, while the VRP-TF spectrum  of bird and pedestrian show relatively concentrated diffuse horizontal stripes. Hence  the VRP-TF spectrum  can be used for LAT recognition.

\subsection{Time-Resolution-Preferred Time-Frequency Spectrum}

Let us set $L_{\rm win}$ to a small value $L_{\rm win,short}$, set $\Delta L_{\rm win} = \Delta L_{\rm win,short}$, and set $ P_{\rm win} =  P_{\rm win,short} > L_{\rm win,short}$. Then we can obtain
the \emph{Time-Resolution-Preferred Time-Frequency Spectrum (TRP-TF spectrum)} of the LAT from the 
$k$-th BS as
\begin{equation}
	\begin{split}
		\begin{aligned}
			\label{deqn_ex1a}
			\!\!
			\mathbf{F}_{k,\rm STFT}^{\rm TRP} \!=\! {\rm STFT} \! \left\{
			\mathbf{Y}_{k,{\rm bin}},
			L_{\rm win,short}, \Delta L_{\rm win,short}, P_{\rm win,short}
			\right\}.
		\end{aligned}
	\end{split}
\end{equation}

Fig.~7  shows the example of the TRP-TF spectrum   observed by BS-A for four categories of LATs. 
It can be seen along the time axis that
the TRP-TF spectrum   displays periodic envelopes caused by the periodic micro-motions of LATs.
The micro-motion frequency of UAV is much higher than that of bird and pedestrian, while the micro-motion frequency of vehicle is also higher than that of bird and pedestrian. 
There are also significant differences 
between the spectrum of UAV and vehicle.
Hence the TRP-TF spectrum   can be effectively used for LAT recognition.

\subsection{Velocity-Transfer Time-Frequency Spectrum}

To further improve the quality of TF spectrum, we consider using matched filtering  to replace the velocity-FFT calculation inside each window, which may improve the velocity resolution without sacrificing the time resolution. 
Meanwhile, note that when the velocity of the target changes periodically, its acceleration will also change periodically. Here we propose to obtain the LAT's 
\emph{Velocity Transfer Time-Frequency Spectrum (VT-TF spectrum)}  composed of the  velocity component's transition amount.

Let us set the window length as $L_{\rm win,transfer}$, set  the sliding step size as $\Delta L_{\rm win,transfer}$, set the minimum and maximum boundary values for velocity matching filtering  as $V_{\rm win,min}$ and $V_{\rm win,max}$, respectively, and set the step size of the velocity matching filtering as $\Delta V_{\rm win}$.
Divide $\mathbf{Y}_{k,{\rm bin}} \in \mathbb{C}^{N\times 1}$ into $G_{\rm transfer} = \lfloor \frac{N-L_{\rm win,transfer}}{\Delta L_{\rm win,transfer}} \rfloor$ groups of signal segment pairs. The $g$-th group signal segment  pair contains two signal vectors $\mathbf{y}_{k,{\rm transfer},g,a} \in \mathbb{C}^{L_{\rm win,transfer}\times 1}$ and $\mathbf{y}_{k,{\rm transfer},g,b} \in \mathbb{C}^{(L_{\rm win,transfer} + 1)\times 1}$,
where 
$\mathbf{y}_{k,{\rm transfer},g,a} = \mathbf{Y}_{k,\rm bin}[(g-1)\Delta L_{\rm win,transfer}+1:(g-1)\Delta L_{\rm win,transfer}+L_{\rm win,transfer}]$
and
$\mathbf{y}_{k,{\rm transfer},g,b} = \mathbf{Y}_{k,\rm bin}[(g-1)\Delta L_{\rm win,transfer}+1:(g-1)\Delta L_{\rm win,transfer}+L_{\rm win,transfer}+1]$.
Then we can construct a velocity search vector as $\mathbf{v}_g = (V_{\rm win,min}:\Delta V_{\rm win}:V_{\rm win,max})^T$, where $(\zeta_1:\Delta \zeta: \zeta_2)^T$ represents constructing a column vector with $\zeta_1$ as the initial value, $\Delta \zeta$ as the step size, and $\zeta_2$ as the termination value. There is ${\rm len}(\mathbf{v}_g) \equiv \frac{V_{\rm win,max}-V_{\rm win,min}+\Delta V_{\rm win}}{\Delta V_{\rm win}} \triangleq I$. 

We refer to $\mathbf{v}_g[i]$ as the $i$-th
\emph{velocity component} of the LAT.
For each $\mathbf{v}_g[i]$, we construct two dictionary vectors $\mathbf{a}_{k,g,a,i} \in \mathbb{C}^{L_{\rm win,transfer}\times 1}$ and $\mathbf{a}_{k,g,b,i} \in \mathbb{C}^{(L_{\rm win,transfer}+1)\times 1}$ that satisfy
\begin{equation}
	\begin{split}
		\begin{aligned}
			\label{deqn_ex1a}
			\!\!\!\!\!\!
\mathbf{a}_{k,g,a,i}[l_a] \!=\!  
e^{-\!j4\pi \frac{f_k \mathbf{v}_g[i] T_r  l_a}{c}},\! \!\! \!\! \quad l_a \! =\! 0,\!...,L_{\rm win,transfer}\!-\!1,
		\end{aligned}
	\end{split}
\end{equation}
\begin{equation}
	\begin{split}
		\begin{aligned}
			\label{deqn_ex1a}
			\!\!\!\!\!\!\!\!\!\!\!\!\!\!\!\!\!\!\!\!\!\!\!\!
			\mathbf{a}_{k,g,b,i}[l_b] \!=\!  
			e^{-\!j4\pi \frac{f_k \mathbf{v}_g[i] T_r  l_b}{c}},\! \!\! \!\! \quad l_b \! =\! 0,\!...,L_{\rm win,transfer},
		\end{aligned}
	\end{split}
\end{equation}
with  $i=1,2,...,I$.
Then we can calculate  two matching values corresponding to $\mathbf{v}_g[i]$ as
\begin{equation}
	\begin{split}
		\begin{aligned}
			\label{deqn_ex1a}
G_{k,g,a,i} = \left| 
\mathbf{a}_{k,g,a,i}^H \mathbf{y}_{k,{\rm transfer},g,a}
\right|,
		\end{aligned}
	\end{split}
\end{equation}
\begin{equation}
	\begin{split}
		\begin{aligned}
			\label{deqn_ex1a}
			G_{k,g,b,i} = \left| 
			\mathbf{a}_{k,g,b,i}^H \mathbf{y}_{k,{\rm transfer},g,b}
			\right|.
		\end{aligned}
	\end{split}
\end{equation}
Define the \emph{velocity transition amount}
on the $i$-th velocity component of the $g$-th signal segment pair for the 
 $k$-th BS  as
\begin{equation}
	\begin{split}
		\begin{aligned}
			\label{deqn_ex1a}
			\!\!\!
			G_{k,g,i} =  \left( 
			\frac{G_{k,g,b,i}}{(L_{\rm win,transfer} \!+\! 1)/L_{\rm win,transfer}}\right)^2
			\! -\! 
			\left(G_{k,g,a,i}\right)^2.
		\end{aligned}
	\end{split}
\end{equation}
Next, we  stack $\{ G_{k,g,i} | i=1,...,I; g = 1,...,G_{\rm transfer}\}$ as a matrix and obtain the VT-TF spectrum of the LAT as $\mathbf{F}_{k}^{\rm VT} \in \mathbb{R}^{I \times G_{\rm transfer}}$, which satisfies
\begin{equation}
	\begin{split}
		\begin{aligned}
			\label{deqn_ex1a}
			\mathbf{F}_{k}^{\rm VT} \ [i,g] =  G_{k,g,i}.
		\end{aligned}
	\end{split}
\end{equation}

\begin{figure*}[!t]
	\centering
	\includegraphics[width=165mm]{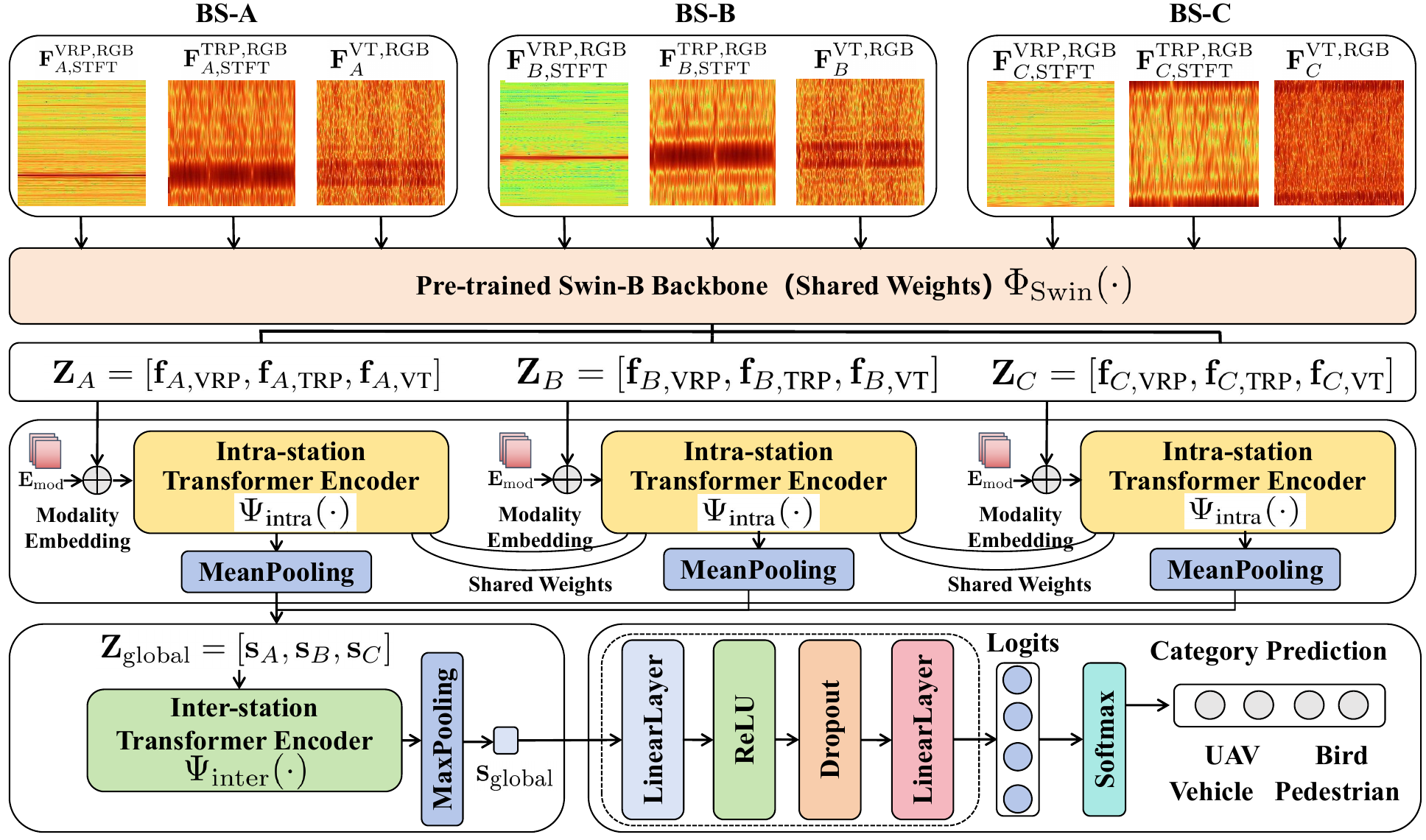}
	\caption{Multi-BS and  multi-scale feature fusion enabled LAT recognition network
		based on Swin-B.}
	\label{fig_1}
\end{figure*}

Fig.~8  shows the examples of the VT-TF spectrum   observed by BS-A for four categories of LATs.
The VT-TF spectrum not only reflects the
micro-motion periodicity of LAT, but also displays more velocity distribution details.
Meanwhile, the VT-TF spectrum is smoother due to the calculation of the transition amount for each velocity component.
It can be seen from Fig.~8 that there are significant differences in the VT-TF spectrum of UAV, bird, vehicle, and 
pedestrian. Hence  the VT-TF spectrum
can be utilized for LAT recognition.

\subsection{Low-Altitude Target Recognition  Network}

We collectively refer to the 
VRP-TF spectrum $\mathbf{F}_{k,\rm STFT}^{\rm VRP}$, 
the TRP-TF spectrum 
$\mathbf{F}_{k,\rm STFT}^{\rm TRP}$,
and the VT-TF spectrum $\mathbf{F}_{k}^{\rm VT}$
 as the \emph{multi-scale feature} of the LAT observed by the $k$-th BS. 
 Then the $k$-th BS can individually visualize
  $\mathbf{F}_{k,\rm STFT}^{\rm VRP}$,
  $\mathbf{F}_{k,\rm STFT}^{\rm TRP}$, 
  $\mathbf{F}_{k}^{\rm VT}$
  as the  RGB images
  $\mathbf{F}_{k,\rm STFT}^{\rm VRP,RGB}$,
  $\mathbf{F}_{k,\rm STFT}^{\rm TRP,RGB}$, 
  $\mathbf{F}_{k}^{\rm VT,RGB}$.
 The dimensions of these RGB images are  
 $3$@$256 \times 256$, in which 
  ``$3$@'' represents the number of channels of the image, and ``$256 \times 256$''  represents the data dimensions in each image channel. 
 All  BSs (BS-A, BS-B, BS-C)  transmit their processed feature images to the data center, thereby forming 
 the
  \emph{multi-BS and multi-scale feature set}, denoted as 
  $\mathcal{F}_{\rm RGB} = \{ 
  \mathbf{F}_{A,\rm STFT}^{\rm VRP,RGB},
 \mathbf{F}_{A,\rm STFT}^{\rm TRP,RGB}, 
 \mathbf{F}_{A}^{\rm VT,RGB}, 
 \mathbf{F}_{B,\rm STFT}^{\rm VRP,RGB},
 \mathbf{F}_{B,\rm STFT}^{\rm TRP,RGB},$ $ 
 \mathbf{F}_{B}^{\rm VT,RGB},
 \mathbf{F}_{C,\rm STFT}^{\rm VRP,RGB},
 \mathbf{F}_{C,\rm STFT}^{\rm TRP,RGB},
 \mathbf{F}_{C}^{\rm VT,RGB}\}$.

  As shown in Fig.~9,
we  design a multi-BS and  multi-scale feature fusion enabled LAT recognition network
 based on swin-transformer-base (Swin-B). 
 Swin-B
is proposed by Microsoft Research and pre-trained on   ImageNet dataset, which  employs  hierarchical structure and  shifted-window based self-attention mechanism, and thus possesses powerful feature extraction capabilities\cite{liu2021swin}.

 We  resize each RGB image in $\mathcal{F}_{\rm RGB}$ to $3$@$224 \times 224$ to match the input requirements of Swin-B.
 Then $\mathbf{F}_{k,\rm STFT}^{\rm VRP,RGB}$, 
 $\mathbf{F}_{k,\rm STFT}^{\rm TRP,RGB}$, and  
 $\mathbf{F}_{k}^{\rm VT,RGB}$
 are 
independently fed into the shared-weight Swin-B backbone 
to extract the corresponding deep features as 
 	\begin{equation}
 		\begin{cases}
 			\mathbf{f}_{k, \text{VRP}} = \Phi_{\text{Swin}}(\mathbf{F}_{k,\rm STFT}^{\rm VRP,RGB}) \\
 			\mathbf{f}_{k, \text{TRP}} = \Phi_{\text{Swin}}(\mathbf{F}_{k,\rm STFT}^{\rm TRP,RGB}) \\
 			\mathbf{f}_{k, \text{VT}} = \Phi_{\text{Swin}}(\mathbf{F}_{k}^{\rm VT,RGB})
 		\end{cases}
 		,
 	\end{equation}
 	where $k \in \{A,B,C\}$, and
 	$\Phi_{\text{Swin}}(\cdot)$ denotes the pre-trained Swin-B feature extractor without  original classification head. 
 
Let us tokenize the extracted feature of the $k$-th BS as 
\begin{equation}
	\mathbf{Z}_{k} = [\mathbf{f}_{k, \text{VRP}}, \mathbf{f}_{k, \text{TRP}}, \mathbf{f}_{k, \text{VT}}].
\end{equation}
Since $\mathbf{f}_{k, \text{VRP}}$,
$\mathbf{f}_{k, \text{TRP}}$,
and $\mathbf{f}_{k, \text{VT}}$
characterize the LAT from distinct resolution perspectives,
 we design a learnable modality token $\mathbf{E}_{\text{mod}}$ to distinguish their intrinsic properties. 
The $\mathbf{E}_{\text{mod}}$ is added to $\mathbf{Z}_k$ to realize modality embedding, and we obtain 
 \begin{equation}
 	\tilde{\mathbf{Z}}_k = \mathbf{Z}_k + \mathbf{E}_{\text{mod}}.
 \end{equation}
 \textcolor{black}{Subsequently, we employ a transformer encoder \cite{vaswani2017attention} (defined as $\Psi_{\text{intra}}(\cdot)$) with multi-head self-attention to capture the complementary relationships among the multi-scale feature within one single BS,} and then 
 we apply mean pooling over the sequence length to obtain
 	the 
 	 comprehensive feature of the $k$-th BS after multi-scale feature fusion as
 	\begin{equation}
 		\mathbf{s}_{k} = \text{MeanPool}(\Psi_{\text{intra}}(\tilde{\mathbf{Z}}_k)).
 \end{equation}
 
 \textcolor{black}{After intra-BS feature interaction, we construct the inter-BS tokenized input sequence as 
 	\begin{equation}
 		\mathbf{Z}_{\rm global} = [\mathbf{s}_{A}, \mathbf{s}_{B}, \mathbf{s}_{C}].
 	\end{equation}
 	Then the aggregated sequence $\mathbf{Z}_{\rm global}$ is fed into another transformer encoder $\Psi_{\text{inter}}(\cdot)$ and processed by max pooling to
 	 obtain the comprehensive feature after multi-BS  fusion as
 	\begin{equation}
 		\mathbf{s}_{\rm global} = \text{MaxPool}(\Psi_{\text{inter}}(\mathbf{Z}_{\rm global})).
 \end{equation}}Finally, the fused global feature $\mathbf{s}_{\rm global}$ is passed through a multi-layer perceptron  classification head to output 
a four-dimensional logit vector corresponding to the predefined LAT categories (UAV, bird, vehicle, pedestrian), which is then  normalized by  Softmax function to realize LAT recognition.

\section{Simulation Results}

In this section, we  evaluate the performance of the proposed LAT recognition scheme.

\vspace{-3mm}

\subsection{Low-Altitude Target Dataset and Partition Protocols}

We construct  CPCs for 10 subtypes of UAVs, 10 subtypes of birds, 10 subtypes of vehicles, and 10 subtypes of pedestrians\footnote{\textcolor{black}{
		The CPCs and more information of these LATs can be found at \url{https://alivn999.github.io/COSMOS-Networked-ISAC-Enabled-Target-Recognition-Towards-Low-Altitude-Economy/}.}}.
The UAVs include 
DJI Mini 2, 
DJI Mini 4 Pro,
DJI Avata,
DJI Mavic 3 Pro,
DJI Mavic 4 Pro,
DJI Inspire 2,
Type-A Firefighting UAV,
Type-B Firefighting UAV,
DJI Matrice 600 Pro, and 
EH 216 S.
The birds include
crow, eagle, kingfisher, parrot, pigeon, ramphastidae, sparrow, seagull, owl, and wild goose.
The vehicles include
AITO M9 2024, Tesla Roadster 2020, Xiaomi SU7 2024,
XPeng G9 2022, ZEEKR 001 2022, Audi A6L 2019,
BYD Song L EV 2024, Geely Galaxy L7 2024,
Leading Ideal L8 2023, and NIO ET5T 2023.
The  pedestrians include 8 walking pedestrians  and 2 running pedestrians.


{To generate the echo signal dataset,}  we set the operating frequency of  ISAC network as $f_0 = 4.9$~GHz, and  set the subcarrier spacing as $\Delta f = 120$ kHz.
Each BS utilizes $M = 256$ subcarriers. 
Meanwhile,  $M_{\rm protect} = 10$ subcarriers are used as frequency protection intervals between adjacent BSs. 
Besides, 
we 
set the array size of TU-UPA as
$N_{T,k}=N^{\parallel}_{T,k}\times N_{T,k}^{\perp}
= 16 \times 4$, and 
set the array size of RU-UPA as
$N_{R,k}=N^{\parallel}_{R,k}\times N_{R,k}^{\perp}= 16 \times 4$.
Moreover,
we set the time interval between adjacent OFDM symbols in the TRF as $T_r = 2.5$~ms, and then use  $N=400$ OFDM symbols for LAT recognition.

Subsequently, 
 we generate $6000$ motion models corresponding to each subtype of LAT by randomly changing their  
paddle frequency, wing frequency, wheel phase,  position, attitude, velocity and other motion  parameters.
Then 
 we  generate a 
LAT echo signal dataset
 containing  $240000$ noiseless echo signal  matrices.
We further extract $240000$ groups of  noiseless symbol domain complex signals of the LAT from the noiseless echo signal  matrices,
in which each group of signals includes the signals from BS-A, BS-B, and BS-C.
Here each group of signals constitutes one target sample.
Next, we add random noise with  signal-to-noise-power-ratio (SNR) 
 belongs to $\{3, 8, 13, 18, 23, \infty \}$~dB to the noiseless symbol domain complex signal, in which the SNRs of multiple BSs in the same target sample are equivalent, and $\infty$ means that no noise is added.
Then  $240000 \times 6 = 1440000$ noisy symbol domain complex signal samples are obtained.

We set 
$L_{\rm win,long}=256$, $\Delta L_{\rm win,long}=1$, $P_{\rm win,long}=256$ to obtain the VRP-TF spectrum $\mathbf{F}_{k,\rm STFT}^{\rm VRP}$ with $k \in \{A,B,C\}$.
We  set 
$L_{\rm win,short}=8$, $\Delta L_{\rm win,short}=1$, $P_{\rm win,short}=1024$ to obtain
the TRP-TF spectrum $\mathbf{F}_{k,\rm STFT}^{\rm TRP}$ with $k \in \{A,B,C\}$. 
 We set  $L_{\rm win,transfer} = 8$, 
$\Delta L_{\rm win,transfer} = 1$
$V_{\rm win,min} = -6.1$,
$V_{\rm win,max} = 6.1$,
$\Delta V_{\rm win} = 0.1$ to obtain  
the VT-TF spectrum $\mathbf{F}_{k}^{\rm VT}$  with $k \in \{A,B,C\}$. 
We visualize these multi-BS and multi-scale features as RGB images, in which  each LAT sample corresponds  to $9$ feature images. 
The image size is set as $3$@$256\times 256$. 
To facilitate the storage and retrieval of the images, we concatenate one  LAT's $9$  feature images    into an aggregated image with a size  $3$@$768\times 768$, which will be re-split into $9$  feature images  during the data loading stage. Finally, we save these  $1440000$ aggregated images  to train, validate, and evaluate the proposed LAT recognition scheme.

\textcolor{black}{
	We consider two data partition protocols. 
	For the \emph{seen-subtype setting}, 
	samples from all target subtypes are randomly divided into training, validation, and testing sets
	 with $1,152,000$, $144,000$, and $144,000$ samples, respectively. 
	For the \emph{unseen-subtype setting}, five subtypes from each LAT category are used for training, validation, and seen-subtype testing, while the remaining five subtypes are reserved for unseen-subtype testing, leading to $576,000$, $72,000$, $72,000$, and $720,000$ samples, respectively. 
	The subtype split is summarized in Table~\ref{tab:seen_unseen_split}.}

\begin{table}[t]
	\centering
	\caption{Subtype split for generalization evaluation.}
	\label{tab:seen_unseen_split}
	\scriptsize
	\begin{tabular}{p{0.12\linewidth}p{0.37\linewidth}p{0.37\linewidth}}
		\toprule
		Category & Seen subtypes & Unseen subtypes \\
		\midrule
		UAV 
		& DJI Mini 2; DJI Avata; DJI Mavic 3 Pro; Type-A Firefighting UAV; DJI Matrice 600 Pro 
		& DJI Mini 4 Pro; DJI Mavic 4 Pro; DJI Inspire 2; Type-B Firefighting UAV; EH 216 S \\
		\midrule
		Bird 
		& Crow; Eagle; Kingfisher; Parrot; Pigeon 
		& Ramphastidae; Sparrow; Seagull; Owl; Wild goose \\
		\midrule
		Vehicle 
		& AITO M9 2024; Tesla Roadster 2020; Xiaomi SU7 2024; XPeng G9 2022; ZEEKR 001 2022 
		& Audi A6L 2019; BYD Song L EV 2024; Geely Galaxy L7 2024; Leading Ideal L8 2023; NIO ET5T 2023 \\
		\midrule
		Pedestrian 
		& Pedestrian 1; Pedestrian 2; Pedestrian 3; Pedestrian 4; Pedestrian 9
		& Pedestrian 5; Pedestrian 6; Pedestrian 7; Pedestrian 8; Pedestrian 10 \\
		\bottomrule
	\end{tabular}
	\normalsize
\end{table}

\begin{table}[t]
	\centering
	\caption{Recognition performance in the seen-subtype setting (\%).}
	\label{tab:seen_subtype_accuracy}
	\footnotesize
	\setlength{\tabcolsep}{1.6pt}
	\begin{tabular}{lccccccc}
		\toprule
		Scheme & 3 dB & 8 dB & 13 dB & 18 dB & 23 dB & $\infty$ & Avg. \\
		\midrule
		BS-A: VRP & 94.70 & 97.69 & 98.79 & 99.00 & 99.14 & 99.11 & 98.07 \\
		BS-A: TRP & 95.51 & 98.50 & 99.20 & 99.43 & 99.39 & 99.44 & 98.58 \\
		BS-A: VT  & 93.01 & 97.63 & 98.88 & 99.16 & 99.30 & 99.32 & 97.88 \\
		\rowcolor{blue!5}
		BS-A: VRP+TRP+VT & 96.52 & 98.88 & 99.42 & 99.57 & 99.63 & 99.65 & 98.94 \\
		\midrule
		BS-A: VRP+TRP+VT & 96.52 & 98.88 & 99.42 & 99.57 & 99.63 & 99.65 & 98.94 \\
		BS-B: VRP+TRP+VT & 97.00 & 98.84 & 99.33 & 99.52 & 99.50 & 99.53 & 98.95 \\
		BS-C: VRP+TRP+VT & 96.75 & 98.81 & 99.38 & 99.52 & 99.53 & 99.55 & 98.92 \\
		\rowcolor{green!7}
		BS-A/B/C: VRP+TRP+VT & 99.52 & 99.91 & 99.95 & 99.97 & 99.97 & 99.97 & 99.88 \\
		\bottomrule
	\end{tabular}
	\normalsize
\end{table}

\begin{table}[t]
	\centering
	\caption{Recognition performance in the unseen-subtype setting (\%).}
	\label{tab:unseen_subtype_accuracy}
	\footnotesize
	\setlength{\tabcolsep}{1.6pt}
	\begin{tabular}{lccccccc}
		\toprule
		Scheme & 3 dB & 8 dB & 13 dB & 18 dB & 23 dB & $\infty$ & Avg. \\
		\midrule
		BS-A: VRP & 87.64 & 90.03 & 91.06 & 91.53 & 91.74 & 92.14 & 90.69 \\
		BS-A: TRP & 90.05 & 93.58 & 94.67 & 95.06 & 95.28 & 95.43 & 94.01 \\
		BS-A: VT  & 87.56 & 92.73 & 94.33 & 94.95 & 95.31 & 95.51 & 93.40 \\
		\rowcolor{blue!5}
		BS-A: VRP+TRP+VT & 91.65 & 94.44 & 95.69 & 96.10 & 96.24 & 96.36 & 95.08 \\
		\midrule
		BS-A: VRP+TRP+VT & 91.65 & 94.44 & 95.69 & 96.10 & 96.24 & 96.36 & 95.08 \\
		BS-B: VRP+TRP+VT & 91.52 & 94.65 & 95.81 & 96.38 & 96.73 & 97.03 & 95.35 \\
		BS-C: VRP+TRP+VT & 92.26 & 94.62 & 95.83 & 96.24 & 96.42 & 96.47 & 95.31 \\
		\rowcolor{green!7}
		BS-A/B/C: VRP+TRP+VT & 96.08 & 97.74 & 98.19 & 98.26 & 98.30 & 98.34 & 97.82 \\
		\bottomrule
	\end{tabular}
	\normalsize
\end{table}

\begin{table}[t]
	\centering
	\caption{Architecture comparison in the unseen-subtype setting (\%).}
	\label{tab:architecture_comparison}
	\footnotesize
	\setlength{\tabcolsep}{1.6pt}
	\begin{tabular}{lccccccc}
		\toprule
		Scheme & 3 dB & 8 dB & 13 dB & 18 dB & 23 dB & $\infty$ & Avg. \\
		\midrule
		Proposed & \cellcolor{green!7}96.08 & \cellcolor{green!7}97.74 & \cellcolor{green!7}98.19 & \cellcolor{green!7}98.26 & \cellcolor{green!7}98.30 & 98.34 & \cellcolor{green!7}97.82 \\
		Swin-B + Mean Fusion & 94.60 & 96.90 & 97.82 & 98.07 & 98.23 & \cellcolor{green!7}98.38 & 97.33 \\
		ConvNeXt-B + Mean Fusion & 93.90 & 95.73 & 96.65 & 97.14 & 97.45 & 97.69 & 96.43 \\
		ViT-B/16 + Mean Fusion & 93.65 & 95.39 & 95.86 & 96.18 & 96.26 & 96.41 & 95.62 \\
		\bottomrule
	\end{tabular}
	\normalsize
\end{table}

\vspace{-3mm}

\subsection{Recognition Performance in the Seen-Subtype Setting}

We  evaluate the recognition performance of the proposed scheme in the seen-subtype setting, where the training, validation, and testing sets contain samples from all target subtypes. 

To investigate the gain brought by multi-scale feature fusion, we take BS-A as an example and compare the schemes using different feature inputs. 
As shown in Table~\ref{tab:seen_subtype_accuracy}, the scheme using VRP+TRP+VT achieves the best performance among the BS-A schemes under all SNR conditions. 
At the low-SNR condition of $3$ dB, the accuracy is improved from $95.51\%$ with the best single feature to $96.52\%$ with multi-scale feature fusion. 
The overall accuracy is also improved from $98.58\%$ to $98.94\%$. 
These results show that the proposed multi-scale features provide complementary target-dependent information for LAT recognition.

We further evaluate the benefit of multi-BS collaboration by comparing the schemes using BS-A, BS-B, BS-C, and all three BSs with the same multi-scale features.  As shown in Table~\ref{tab:seen_subtype_accuracy},
the multi-BS scheme consistently outperforms the single-BS schemes across all SNR levels. 
Compared with the best single-BS, the BS-A/B/C scheme improves the accuracy from $97.00\%$ to $99.52\%$ at $3$ dB, and from $98.95\%$ to $99.88\%$ in terms of overall accuracy. 
These results verify that multi-BS collaboration further exploits complementary observations from distributed BSs and brings additional recognition gains.

\vspace{-3mm}

\subsection{Generalization Performance in the Unseen-Subtype Setting}

We further evaluate the generalization performance of the proposed scheme in the unseen-subtype setting, where the subtypes used for testing are not observed during training.

As shown in Table~\ref{tab:unseen_subtype_accuracy}, multi-scale feature fusion improves the recognition accuracy of BS-A under all SNR conditions. 
At the low-SNR condition of $3$ dB, the accuracy is improved from $90.05\%$ with the best single feature to $91.65\%$ with VRP+TRP+VT. 
The overall accuracy is also improved from $94.01\%$ to $95.08\%$. 
These results indicate that the proposed multi-scale features provide complementary information that remains effective for unseen target subtypes.

Moreover, multi-BS collaboration further improves the unseen-subtype recognition performance. 
Compared with the best single-BS  scheme, the BS-A/B/C scheme improves the accuracy from $92.26\%$ to $96.08\%$ at $3$ dB, and from $95.35\%$ to $97.82\%$ in terms of overall accuracy. 
This demonstrates that distributed BS observations provide complementary aspect-dependent information, thereby enhancing the generalization ability of LAT recognition.

\vspace{-3mm}

\subsection{Comparison with Different Network Architectures}

We further compare the proposed scheme with several representative network architectures in the unseen-subtype setting. 
For a fair comparison, all the compared schemes use the same training and testing data partition. 
The baseline schemes adopt simple mean fusion to aggregate the deep features extracted from different BSs and different time-frequency representations, while replacing the backbone with Swin-B, ConvNeXt-B\cite{9879745}, and ViT-B/16\cite{dosovitskiy2021imageworth16x16words}, respectively.

As shown in Table~\ref{tab:architecture_comparison}, the proposed scheme achieves higher accuracy than the simple mean fusion baselines under most SNR conditions and obtains the best overall accuracy. 
Compared with Swin-B + Mean Fusion, which uses the same backbone but a simpler fusion strategy, the proposed scheme improves the accuracy from $94.60\%$ to $96.08\%$ at $3$ dB and from $97.33\%$ to $97.82\%$ in terms of overall accuracy. 
This indicates that explicitly modeling intra-BS feature interactions and inter-BS collaborative interactions is more effective than directly averaging the extracted features.

Among the simple mean fusion baselines, Swin-B achieves better overall performance than ConvNeXt-B and ViT-B/16. 
This suggests that the shifted-window attention mechanism is suitable for extracting discriminative representations from the visualized time-frequency features. 
Nevertheless, the main performance gain of the proposed scheme comes from the hierarchical fusion of multi-scale and multi-BS observations rather than from simply increasing the backbone complexity.

\section{Conclusions}

In this paper,
we  proposed a
multi-BS collaboration and multi-scale feature fusion
enabled 
LAT recognition scheme   for ISAC  network. 
We  formulated the motion equations and echo signals for UAV, bird, 
vehicle, and pedestrian under networked ISAC   scenario.
We extracted the 
VRP-TF spectrum,  TRP-TF spectrum, and  VT-TF spectrum observed by each BS from the echo signals, which was  referred to as multi-scale feature. 
Next, we designed a multi-BS and multi-scale feature fusion enabled LAT recognition network with Swin-B, 
which 
employed the visualized images of multi-scale feature
to jointly recognize the target.
We generated a massive echo signal dataset comprising 1,440,000 samples for LAT recognition within ISAC network. This dataset can serve as a public benchmark to evaluate our proposed scheme and facilitate future research.
Simulation results demonstrate that the proposed scheme realizes high recognition accuracy and robust unseen-subtype generalization, confirming the effectiveness of multi-scale feature fusion and  multi-BS collaboration.

\bibliographystyle{ieeetr}
\bibliography{ref.bib}

@misc{dosovitskiy2021imageworth16x16words,
      title={An Image is Worth 16x16 Words: Transformers for Image Recognition at Scale}, 
      author={Alexey Dosovitskiy and Lucas Beyer and Alexander Kolesnikov and Dirk Weissenborn and Xiaohua Zhai and Thomas Unterthiner and Mostafa Dehghani and Matthias Minderer and Georg Heigold and Sylvain Gelly and Jakob Uszkoreit and Neil Houlsby},
      year={2021},
      eprint={2010.11929},
      archivePrefix={arXiv},
      primaryClass={cs.CV},
      url={https://arxiv.org/abs/2010.11929}, 
}

@INPROCEEDINGS{9879745,
  author={Liu, Zhuang and Mao, Hanzi and Wu, Chao-Yuan and Feichtenhofer, Christoph and Darrell, Trevor and Xie, Saining},
  booktitle={2022 IEEE/CVF Conference on Computer Vision and Pattern Recognition (CVPR)}, 
  title={A ConvNet for the 2020s}, 
  year={2022},
  volume={},
  number={},
  pages={11966-11976},
  doi={10.1109/CVPR52688.2022.01167}}

@article{vaswani2017attention,
  title={Attention is all you need},
  author={Vaswani, Ashish and Shazeer, Noam and Parmar, Niki and Uszkoreit, Jakob and Jones, Llion and Gomez, Aidan N and Kaiser, {\L}ukasz and Polosukhin, Illia},
  journal={Advances in neural information processing systems},
  volume={30},
  year={2017}
}

@ARTICLE{11250835,
  author={Luo, Hongliang and Zhang, Tengyu and Zhao, Chuanbin and Wang, Yucong and Lin, Bo and Jiang, Yuhua and Luo, Dongqi and Gao, Feifei},
  journal={IEEE Wireless Commun.}, 
  title={Integrated Sensing and Communications Framework for {6G} Networks}, 
  year={2025},
  volume={32},
  number={6},
  pages={102-109},
  doi={10.1109/MWC.2025.3599658}}

@inproceedings{liu2021swin,
  title={Swin transformer: Hierarchical vision transformer using shifted windows},
  author={Liu, Ze and Lin, Yutong and Cao, Yue and Hu, Han and Wei, Yixuan and Zhang, Zheng and Lin, Stephen and Guo, Baining},
  booktitle={Proceedings of the IEEE/CVF international conference on computer vision},
  pages={10012--10022},
  year={2021}
}

@article{gabor1946theory,
  title={Theory of communication. Part 1: The analysis of information},
  author={Gabor, Dennis},
  journal={Journal of the Institution of Electrical Engineers-part III: radio and communication engineering},
  volume={93},
  number={26},
  pages={429--441},
  year={1946},
  publisher={IET}
}

@ARTICLE{1455039,
  author={Allen, J.B. and Rabiner, L.R.},
  journal={Proceedings of the IEEE}, 
  title={A unified approach to short-time Fourier analysis and synthesis}, 
  year={1977},
  volume={65},
  number={11},
  pages={1558-1564},
  doi={10.1109/PROC.1977.10770}}

@misc{RN16,
  author = {Huaweicloud},
  year = {Available on March 13, 2026},
  url = {https://developer.huaweicloud.com/competition/information/1300000110},
  urldate = {March 13, 2026},
  title = {The 5th Wireless Big Data Competition, Accessed: https://developer.huaweicloud.com/competition/information/1300000110},
  urldate = {March 13, 2026}
}

@misc{CMUmocap,
  author       = {{Carnegie Mellon University (CMU)}},
year = {Available on March 13, 2026},
  title        = {The Motion Research Laboratory, Accessed: http://mocap.cs.cmu.edu},
  url          = {http://mocap.cs.cmu.edu},
  urldate      = {2026-03-13}, 
  organization = {Motion Research Laboratory},
  note         = {Accessed: http://mocap.cs.cmu.edu, 2026-03-13}
}

@ARTICLE{11159257,
  author={Luo, Hongliang and Chu, Zhonghua and Zhang, Tengyu and Zhao, Chuanbin and Lin, Bo and Gao, Feifei},
  journal={IEEE J. Sel. Areas Commun.}, 
  title={AirGuard: {UAV} and Bird Recognition Scheme for Integrated Sensing and Communications System}, 
  year={2026},
  volume={44},
  number={},
  pages={835-848},
  doi={10.1109/JSAC.2025.3608760}}

@INPROCEEDINGS{11069481,
  author={Xue, Junyao and Zhang, Qixun and Ma, Dingyou and Wei, Jiachen},
  booktitle={2025 10th International Conference on Computer and Communication System (ICCCS)}, 
  title={DC-Former Network Empowered UAV and Bird Recognition Based on Integrated Sensing and Communication System}, 
  year={2025},
  volume={},
  number={},
  pages={927-932},
  doi={10.1109/ICCCS65393.2025.11069481}}

@ARTICLE{11077832,
  author={Wei, Jiachen and Ma, Dingyou and He, Feiyang and Zhang, Qixun and Feng, Zhiyong and Liu, Zhengfeng and Liang, Taohong},
  journal={IEEE Trans. Wireless Commun.}, 
  title={UAV’s Rotor Micro-Doppler Feature Extraction Using Integrated Sensing and Communication Signal: Algorithm Design and Testbed Evaluation}, 
  year={2025},
  volume={24},
  number={12},
  pages={10166-10182},
  doi={10.1109/TWC.2025.3578033}}

@ARTICLE{11151696,
  author={Huang, Yixuan and Yang, Jie and Xia, Shuqiang and Wen, Chao-Kai and Jin, Shi},
  journal={IEEE Trans. Wireless Commun.}, 
  title={Learned Off-Grid Imager for Low-Altitude Economy With Cooperative ISAC Network}, 
  year={2026},
  volume={25},
  number={},
  pages={3333-3348},
  doi={10.1109/TWC.2025.3603255}}

@ARTICLE{10226276,
  author={Wei, Zhiqing and Xu, Ruizhong and Feng, Zhiyong and Wu, Huici and Zhang, Ning and Jiang, Wangjun and Yang, Xiaoyu},
  journal={IEEE Trans. Veh. Technol.}, 
  title={Symbol-Level Integrated Sensing and Communication Enabled Multiple Base Stations Cooperative Sensing}, 
  year={2024},
  volume={73},
  number={1},
  pages={724-738},
  doi={10.1109/TVT.2023.3304856}}

@ARTICLE{11134125,
  author={Yan, Shaoqiang and Luo, Hongliang and Yang, Ping and Zhao, Jianwei and Gao, Feifei},
  journal={IEEE Trans. Wireless Commun.}, 
  title={{UAV} Trajectory Monitoring for Integrated Sensing and Communications System}, 
  year={2026},
  volume={25},
  number={},
  pages={2733-2747},
  doi={10.1109/TWC.2025.3598799}}

@INPROCEEDINGS{11162253,
  author={Zhang, Ruihang and Xue, Jiayin and Zhang, Tingting},
  booktitle={2025 IEEE International Conference on Communications Workshops (ICC Workshops)}, 
  title={Reliable Clutter Suppression for Slow-Moving Weak Target Radar Detection}, 
  year={2025},
  volume={},
  number={},
  pages={354-359},
  doi={10.1109/ICCWorkshops67674.2025.11162253}}

@article{XIANG2024100140,
title = {Autonomous {eVTOL}: A summary of researches and challenges},
journal = {Green Energy Intell. Transp.},
volume = {3},
number = {1},
pages = {100140},
year = {Feb. 2024},
issn = {2773-1537},
doi = {https://doi.org/10.1016/j.geits.2023.100140},
url = {https://www.sciencedirect.com/science/article/pii/S2773153723000762},
author = {Senwei Xiang and Anhuan Xie and Minxiang Ye and Xufei Yan and Xiaojia Han and Hongjiao Niu and Qiang Li and Haishan Huang}
}

@INPROCEEDINGS{10625724,
  author={Ma, Dingyou and others},
  booktitle={Proc. 2nd Int. Conf. Mobile Internet, Cloud Comput. Inf. Security}, 
  title={Performance Evaluation of Micro-{Doppler} Based {UAV} Identification Using Different {5G} Frame Structures}, 
  year={Changsha City, China, Apr. 2024, pp. 173--179},
  volume={},
  number={},
  doi={10.1109/MICCIS63508.2024.00037}}

@article{dai2015euler,
  title={Euler--Rodrigues formula variations, quaternion conjugation and intrinsic connections},
  author={Dai, Jian S},
  journal={Mech. Mach. Theory},
  volume={92},
  pages={144--152},
  year={Oct. 2015},
  publisher={Elsevier}
}

@inproceedings {mesh1, 
booktitle = {Proc. Eurographics Italian Chapter Conf.}, 
title = {{MeshLab: an open-source mesh processing tool}}, 
author = {Cignoni, Paolo and others}, 
year = {Jul. 2008}, 
publisher = {The Eurographics Association}, ISBN = {978-3-905673-68-5}, 
DOI = {10.2312/LocalChapterEvents/ItalChap/ItalianChapConf2008/129-136} }

@ARTICLE{8675384,
  author={Fotouhi, Azade and others},
  journal={IEEE Commun. Surv. Tutor.}, 
  title={Survey on {UAV} Cellular Communications: Practical Aspects, Standardization Advancements, Regulation, and Security Challenges}, 
  year={Mar. 2019},
  volume={21},
  number={4},
  pages={3417-3442},
  doi={10.1109/COMST.2019.2906228}}

@ARTICLE{10077453,
  author={Javaid, Shumaila and Saeed, Nasir and Qadir, Zakria and Fahim, Hamza and He, Bin and Song, Houbing and Bilal, Muhammad},
  journal={IEEE Trans. Intell. Trans. Sys.}, 
  title={Communication and Control in Collaborative {UAVs:} Recent Advances and Future Trends}, 
  year={Jun. 2023},
  volume={24},
  number={6},
  pages={5719-5739},
  doi={10.1109/TITS.2023.3248841}}

@Article{drones3010013,
AUTHOR = {Alsamhi, Saeed H. and others},
TITLE = {Collaboration of Drone and Internet of Public Safety Things in Smart Cities: An Overview of {QoS} and Network Performance Optimization},
JOURNAL = {Drones},
VOLUME = {3},
YEAR = {Jan. 2019},
NUMBER = {1},
ARTICLE-NUMBER = {13}
}

@ARTICLE{9666755,
  author={Hoque, Mohammad Aminul and others},
  journal={IEEE Internet Things J.}, 
  title={{IoTaaS:} Drone-Based Internet of Things as a Service Framework for Smart Cities}, 
  year={Jul. 2022},
  volume={9},
  number={14},
  pages={12425-12439},
  doi={10.1109/JIOT.2021.3137362}}

@ARTICLE{8795473,
  author={Alsamhi, Saeed H. and Ma, Ou and Ansari, Mohammad Samar and Almalki, Faris A.},
  journal={IEEE Access}, 
  title={Survey on Collaborative Smart Drones and Internet of Things for Improving Smartness of Smart Cities}, 
  year={Aug. 2019},
  volume={7},
  number={},
  pages={128125-128152},
  doi={10.1109/ACCESS.2019.2934998}}

@article{HUANG2024100694,
title = {Low-altitude intelligent transportation: System architecture, infrastructure, and key technologies},
journal = {J. Ind. Inf. Integr.},
volume = {42},
pages = {100694},
year = {Nov. 2024},
issn = {2452-414X},
doi = {https://doi.org/10.1016/j.jii.2024.100694},
url = {https://www.sciencedirect.com/science/article/pii/S2452414X24001377},
author = {Changqing Huang and Shifeng Fang and Hua Wu and Yong Wang and Yichen Yang}
}

@ARTICLE{10608169,
  author={Zheng, Bowen and Liu, Fan},
  journal={IEEE Wireless Commun. Lett.}, 
  title={Random Signal Design for Joint Communication and {SAR} Imaging Towards Low-Altitude Economy}, 
  year={Oct. 2024},
  volume={13},
  number={10},
  pages={2662-2666},
  doi={10.1109/LWC.2024.3432892}}

@ARTICLE{10723207,
  author={Huang, Hailong and Su, Jiangcheng and Wang, Fei-Yue},
  journal={IEEE Trans. Intell. Veh.}, 
  title={The Potential of Low-Altitude Airspace: The Future of Urban Air Transportation}, 
  year={Aug. 2024},
  volume={9},
  number={8},
  pages={5250-5254},
  doi={10.1109/TIV.2024.3483889}}

@ARTICLE{2023arXiv231109047J,
       author = {{Jiang}, Yihang and others},
        title = "{{6G} non-terrestrial networks enabled low-altitude economy: Opportunities and challenges}",
      journal = {arXiv e-prints},
         year = 2023,
        month = nov,
          eid = {arXiv:2311.09047},
        pages = {arXiv:2311.09047},
          doi = {10.48550/arXiv.2311.09047},
archivePrefix = {arXiv},
       eprint = {2311.09047},
 primaryClass = {cs.IT},
       adsurl = {https://ui.adsabs.harvard.edu/abs/2023arXiv231109047J}
}

@ARTICLE{2024arXiv240412705L,
       author = {{Lu}, Xi and {Wei}, Zhiqing and {Xu}, Ruizhong and {Wang}, Lin and {Lu}, Bohao and {Piao}, Jinghui},
        title = "{Integrated sensing and communication enabled multiple base stations cooperative {UAV} detection}",
      journal = {arXiv e-prints},
         year = 2024,
        month = apr,
          eid = {arXiv:2404.12705},
        pages = {arXiv:2404.12705},
          doi = {10.48550/arXiv.2404.12705},
archivePrefix = {arXiv},
       eprint = {2404.12705},
 primaryClass = {eess.SP},
       adsurl = {https://ui.adsabs.harvard.edu/abs/2024arXiv240412705L}
}

@ARTICLE{7523373,
  author={Shafin, Rubayet and Liu, Lingjia and Zhang, Jianzhong and Wu, Yik-Chung},
  journal={IEEE Trans. Wireless Commun.}, 
  title={{DoA} Estimation and Capacity Analysis for {3-D} Millimeter Wave Massive-{MIMO/FD-MIMO} {OFDM} Systems}, 
  year={Oct. 2016},
  volume={15},
  number={10},
  pages={6963-6978},
  doi={10.1109/TWC.2016.2594173}}

@ARTICLE{itu,
       author = {ITU},
        title = "{Future technology trends of terrestrial international mobile telecommunications systems towards 2030 and beyond}",
         year = {2022}
}

@ARTICLE{9606831,
  author={Cui, Yuanhao and Liu, Fan and Jing, Xiaojun and Mu, Junsheng},
  journal={IEEE Netw.}, 
  title={Integrating Sensing and Communications for Ubiquitous {IoT}: Applications, Trends, and Challenges}, 
  year={Nov. 2021},
  volume={35},
  number={5},
  pages={158-167},
  doi={10.1109/MNET.010.2100152}}

@ARTICLE{202310141,
  author={Liu, Fan and Cui, Yuanhao and Masouros, Christos and Xu, Jie and Han, Tony Xiao and Eldar, Yonina C. and Buzzi, Stefano},
  journal={IEEE J. Sel. Areas Commun.}, 
  title={Integrated Sensing and Communications: Toward Dual-Functional Wireless Networks for 6{G} and Beyond}, 
  year={Jun. 2022},
  volume={40},
  number={6},
  pages={1728-1767},
  doi={10.1109/JSAC.2022.3156632}}

@ARTICLE{9947033,
  author={Du, Zhen and Liu, Fan and Yuan, Weijie and Masouros, Christos and Zhang, Zenghui and Xia, Shuqiang and Caire, Giuseppe},
  journal={IEEE Trans. Wireless Commun.}, 
  title={Integrated Sensing and Communications for {V2I} Networks: Dynamic Predictive Beamforming for Extended Vehicle Targets}, 
  year={Jun. 2023},
  volume={22},
  number={6},
  pages={3612-3627},
  doi={10.1109/TWC.2022.3219890}}

@ARTICLE{9898900,
  author={Gao, Zhen and Wan, Ziwei and Zheng, Dezhi and Tan, Shufeng and Masouros, Christos and Ng, Derrick Wing Kwan and Chen, Sheng},
  journal={IEEE Trans. Wireless Commun.}, 
  title={Integrated Sensing and Communication With {mmWave} Massive {MIMO}: A Compressed Sampling Perspective}, 
  year={Mar. 2023},
  volume={22},
  number={3},
  pages={1745-1762},
  doi={10.1109/TWC.2022.3206614}}

@ARTICLE{9040264,  author={Giordani, Marco and Polese, Michele and Mezzavilla, Marco and Rangan, Sundeep and Zorzi, Michele},  journal={IEEE Commun. Mag.},   title={Toward 6{G} Networks: Use Cases and Technologies},   year={Mar. 2020},  volume={58},  number={3},  pages={55-61},  doi={10.1109/MCOM.001.1900411}}

@ARTICLE{9573459,  author={Wang, Wei and Zhang, Wei},  journal={IEEE Trans. Wireless Commun.},   title={Jittering Effects Analysis and Beam Training Design for {UAV} Millimeter Wave Communications},   year={Oct. 2022},  volume={21},  number={5},  pages={3131-3146},  doi={10.1109/TWC.2021.3118558}}

\vfill

\end{document}